\documentstyle[12pt]{article}

\begin{document}
\input{psfig.sty}
\begin{flushright}
\baselineskip=12pt
MADPH-99-1133 \\
\end{flushright}

\begin{center}
\vglue 1.5cm
{\Large\bf Scale, Gauge Couplings,
  Soft Terms and Toy Compactification in M-theory on $S^1/Z_2$\\}
\vglue 2.0cm
{\Large  Tianjun Li~\footnote{E-mail: li@pheno.physics.wisc.edu,
phone: (608) 262-9820, fax: (608) 262-8628.}}
\vglue 1cm
\begin{flushleft}
Department of Physics, University of Wisconsin, Madison, WI 53706,
U. S. A.
\end{flushleft}
\end{center}

\vglue 1.5cm
\begin{abstract}
In M-theory on $S^1/Z_2$, we point out
that to be consistant, we should keep the scale, gauge couplings
and soft terms at next 
order, and obtain
the soft term relations: $M_{1/2} = -A$,
$|{{M_{0}}/{M_{1/2}}}| \leq 
{1/{\sqrt 3}}$ in the
standard embedding and $M_{1/2}=-A$ in the non-standard embedding
with five branes and $K_{5,n}=0$.
 We construct a toy compactification model which 
includes higher order
terms in 4-dimensional Lagrangian in standard 
embedding, and discuss its scale, gauge couplings, soft terms,
and show that the higher order terms do
affect the scale, gauge couplings and especially the soft terms
if the next order correction was not small.
We also construct a toy compactification
model in non-standard embedding with five branes
and discuss its phenomenology. We argue that one might not
push the physical Calabi-Yau manifold's volume to zero at
any point along the eleventh dimension. 
\\[1ex]
PACS: 11.25.Mj; 04.65.+e; 11.30.Pb; 12.60. Jv
\\[1ex]
Keywords: M-theory; Compactification; Supersymmetry; Scale;
Coupling

\end{abstract}

\vspace{0.5cm}
\begin{flushleft}
\baselineskip=12pt
August 1999\\
\end{flushleft}
\newpage
\setcounter{page}{1}
\pagestyle{plain}
\baselineskip=14pt

\section{Introduction}

M-theory on $S^1/Z_2$ suggested by Horava and 
Witten~\cite{HW} 
is a 11-dimensional Supergravity theory with two boundaries where the
two $E_8$  Yang-Mills fields live on respectively. A lot of study on
compactification ( standard embedding and non-standard embedding ) and 
phenomenology have been done~\cite{ Witten,Horava,BD,KC,AQ,AQR,EDCJ,VK,LLN,TIAN,BL,
JUN,JUNJUN,HLYLZH, NOY,YNOY, LT,LOD,LODW, LOW,CKM,BKL,EFN,DM,KBENAKLI,
KARIM,KARIMB,SST,NSLPT,NSLOW, LOSW, JELPP, DVN, CIMUNOZ, JEDVN,
DLOW, TKJKHS, DGCMUNOZ, BGVKL, JONELLIS}. And as we know, the key which connects the
M-theory on $S^1/Z_2$ to the low energy phenomenology is the SUSY breaking
soft terms, because if we knew the soft terms, by running RGE, we can
obtain the low energy SUSY particle spectra and discuss their
search in the future collider. On the other hand, if we knew the
low energy SUSY particle spectra from future collider, we might derive 
the soft terms by solving RGE~\cite{MRAMOND}.
 Therefore, we need to understand
the SUSY breaking soft terms as well as possible.

The orginal eleven-dimensional Lagrangian obtained by Horava and 
Witten~\cite{HW} is at order of $\kappa^{2/3}$. And
Witten's solution to
the compactification of M-theory on $S^1/Z_2$ which have N=1 supersymmetry in
4-dimension just considered the next order expansion of the
metric~\cite{Witten}.
Moreover, after compactifying the
theory on the deformed Calabi-Yau manifold, 
A. Lukas, B. A. Ovrut and D. Waldram obtained
the 4-dimensional Lagrangian 
which  is at order of $\kappa^{4/3}$~\cite{LOD}. The explicit expansion
parameter which is defined in the following section, is $x$.
However, the soft terms: $M_{1/2}, M_0^2, A$ which were obtained previously
 in standard embedding for the simplest compactification
~\cite{CKM},
include the higher order correction, i.e., when 
we expand the soft term $M_{1/2}, M_0^2, A$ as polynomial in x for $ |x| < 1$,
we have the terms which are proportional to $x^n$ where $ n > 1$. This is
 inconsistent to
the original calculation, because we did not include higher order
correction to the previous Lagrangian, or the K\"ahler potential, gauge
kinetic function and superpotential. In short,
the correct soft terms should be at order of $x$, and 
this is a good approximation only when $x$ is small. In this
paper, we calculate the soft terms  $M_{1/2}, M_0^2, A$  at order
of $x$ in standard embedding, 
and compare with previous calculation. We find out that
$ x < 0.2$, we have good approximation, we will have
observable deviation if $x$
is large, i. e., $x > 0.5$. Therefore,the phenomenology discussions,
 like low energy SUSY particle spectra
and collider serach, should keep in the small x region.
In addition, we also discuss the scale and
gauge couplings at order $x$, and the detail discussions 
are similar to those in
~\cite{JUNJUN, HLYLZH,
DGCMUNOZ} with small $x$. Furthermore, we obtain the follwoing
soft term relations in standard embedding
or non-standard embedding without five brane for the simplest 
compactification:
\begin{eqnarray}
M_{1/2} = -A ~,~
|{{M_{0}}\over\displaystyle {M_{1/2}}}| \leq 
{1\over\displaystyle \sqrt 3}
 ~.~\,
\end{eqnarray}

Because the orginal eleven-dimensional Lagrangian  is at order of $\kappa^{2/3}$. 
And Witten's solution is at next order, it is very difficult to
do the calculation which includes higher order correction
 unless one makes some
simplification. In order to discuss higher order correction
and compare the differences of the
 soft terms, we use the following 
 ansatz in our toy model:
(1) we just consider the Lagrangian which was obtained by Horava and
Witten~\cite{HW},  (2) we do not consider the higher order expansion of 
the deformed metric, (3) we do not
consider the higher order correction to the bulk fields. 
Because the massive modes' contributions   
are supressed by the factor ${V_p^{1/6} \over\displaystyle {\pi \rho_p}}$
which is at about 0.1 order and very small
 when we consider the 
intermediate unification~\cite{JUNJUN, DGCMUNOZ},
we just consider the massless modes.

In standard embedding,
under this ansatz and considering the Calabi-Yau manifold with
Hodge-Betti number $h_{(1,1)}=1, h_{(2, 1)}=0$, 
we  calculate the 
higher order correction to the Lagrangian, or the scale, gauge couplings,
 K\"ahler potential,  gauge 
kinetic function and superpotential. Under $x^n =0$ limit for $n > 1$,
our results are the same as those obtained previously~\cite{NOY, LOD}.
In this toy model compactification,  
we find out that: 

(I) In order to
keep $g_{11, 11}$ positive,
 we can not push the physical Calabi-Yau manifold's volume
in the hidden sector to zero.

(II) $x=1$ is not the M-theory limit.  $x=3$ or $ y=1 $ 
( where y is one-third of x ) is the limit
which pushes the physical
 Calabi-Yau manifold's volume in the hidden sector 
to zero,
but, in order to keep $g_{11, 11}$ positive, we require
that $x < 3/2$ or $y < 1/2$. And 
$x = 3/2$ or $y = 1/2$ is considered as M-thoery limit in
this paper. In other words, we define the M-theory limit as the 
limit which
keeps the signature of the metric invariant, i. e., the
total high order corrections to the metric are less than
the zeroth order metric in each component of the metric.

(III) We have strong contraints on the gauge coupling 
$\alpha_H$ and GUT scale $M_H$ in the
hidden sector:
\begin{eqnarray}
{1 \over {27}} \alpha_{GUT} < \alpha_H < 27 \alpha_{GUT}
~,~\, 
\end{eqnarray}
\begin{eqnarray}
{1\over \sqrt 3} M_{GUT} < M_H < \sqrt 3 M_{GUT}
~.~\, 
\end{eqnarray}

(IV) We notice that, if y is small ( $ y < 0.2$ ), in large 
parameter space, 
the magnitude of the gaugino mass
is larger than that of the scalar mass, and if y is large ( $ y > 0.3$ ), 
in the most of the parameter space, the magnitude of the gaugino mass
is smaller than that of the scalar mass. However, in the previous 
soft term analysis~\cite{CKM}, the magnitude of the
gaugino mass is often larger
than that of the scalar mass in the standard embedding. 
In addition, we compare the soft term deviations with the two scenarios
discussed in the second paragrah.

Non-standard embedding in M-theory on $S^1/Z_2$ is also an interesting
subject~\cite{KARIM, NSLPT, NSLOW}. 
We can embed the spin connection to the two $E_8$ gauge
fields, so, after compactification, the
gauge groups in the observable sector and  hidden sector
will be $G^O$ and $G^H$, which are subgroups of
$E_8$.  And we can include the five branes, states which are
essentially non-perturbative in heterotic string.
The presence of the five branes is very important to
the three generation GUT model building because they
introduce more freedom in the anomaly cancellation 
condition that consistent vacvua need to satisfy
~\cite{NSLOW, DLOW}. And 
introducing five branes will affect the gauge kinetic function, K\"ahler
potential and non-perturbative superpotential by gaugino condensation 
in the next order because
one introduces more moduli whose real parts are the positions of
the five-branes.
Without five brane, the non-standard embedding's
results are similar to those in the standard embedding. 
With five branes,
we discuss the scale, gauge couplings and
soft terms  to the next order,
and we find that if there are no
K\"ahler potential for the five brane moduli ( $K_5$ ),
we obtain the soft term relation:
\begin{eqnarray}
M_{1/2} = -A 
 ~.~\,
\end{eqnarray} 
This may be interesting in the low energy phenomenology analysis.
Furthermore, we 
  consider 
the toy model compactification and calculate
its K\"ahler potential, gauge kinetic function, superpotential, and
discuss its scale, gauge couplings and soft terms.
Our results are:

(I) In order to
keep $g_{11, 11}$ positive,
 we can not push the physical Calabi-Yau manifold's volume
in the hidden sector to zero.

(II) M-theory limit is:   $  y_{5b} = {1\over 2}$
or $y_{5b} = - 1$, where $y_{5b} $ is defined in the following section.
In order to keep $ g_{11, 11} $ positive, we require that
$  y_{5b} < {1\over 2}$, and in order to keep the signature of 
metric $g_{\mu \nu}$ invariant, we require that
 $y_{5b} > - 1$. In short, we obtain:
 $ -1 <  y_{5b} < {1\over 2}$. 
 
(III) We have strong contraints on the gauge coupling 
$\alpha_H$ and GUT scale $M_H$ in the
hidden sector:
\begin{eqnarray}
{1 \over {64}} \alpha_{GUT} < \alpha_H < 64 \alpha_{GUT}
~,~\, 
\end{eqnarray}
\begin{eqnarray}
{1\over 2} M_{GUT} < M_H < 2 M_{GUT}
~.~\, 
\end{eqnarray}

Of course, we have
 more freedom in phenomenology discussion if we include five branes,
 for we introduce the new parameters $\beta^i$ where
 i=1, N. Therefore, we do not do the numerical analysis of the soft terms
 because we just have 
 three soft term parameters: $M_{1/2}, M_0^2$, and $A$, and  we can make them
 as free parameters by varying  $\beta^i$ and $\epsilon$ which is
 defined in the following section.   

Finally, we argue that, in general, we might not
push the physical Calabi-Yau manifold's volume to zero at
any point along the eleventh dimension. 

Our conventions are the following. 
We denote the eleven-dimensional coordinates
by $x^0,\ldots,x^9,x^{11}$ and the corresponding indices by
$I,J,K,\ldots=0,\ldots,9,11$. The orbifold  $S^1/Z_2$ is chosen in the
$x^{11}$--direction, so we assume that  $x^{11}\in [-\pi \rho  ,\pi\rho ]$ with
the endpoints identified as $x^{11}\sim x^{11}+2\pi \rho$. The $Z_2$ symmetry
acts as $x^{11}\rightarrow -x^{11}$. Then, there exist two ten-dimensional
hyperplanes, $M^{10}_i$ with $i=1,2$, locally specified by the
conditions $x^{11}=0$ and $x^{11}=\pi\rho$, which are fixed under the
action of the $Z_2$ symmetry. When we compactify the theory on a
Calabi-Yau three-fold, we will use indices $A,B,C,\ldots=4,\ldots,9$
for the Calabi-Yau coordinates, and indices $\mu,\nu\ldots=0,\ldots,3$
for the coordinates of the remaining, uncompactified, four-dimensional
space. Holomorphic and antiholomorphic coordinates on the Calabi-Yau space 
will be labeled by $a,b,c,\ldots$ and $\bar a,\bar b,\bar c,\ldots$.
In addition, we denote the hyperplane $M_1^{10}$ as the observable sector,
and the hyperplane $M_2^{10}$ as the hidden sector.
 
\section{Scale, Gauge Couplings,  Soft Terms and Toy Compactification in Standard Embedding}

\subsection{ Scale,Gauge Couplings, K\"ahler Function and Soft Terms Revisit}

First, let us  review the gauge couplings, gravitational
coupling and the physical eleventh dimension
 radius in the M-theory~\cite{HLYLZH}. The relevant
11-dimensional Lagrangian is given by~\cite{HW}
\begin{eqnarray}
L_B&=&-{1\over {2 \kappa^2}}\int_{M^{11}}d^{11}x\sqrt g
R - \sum_{i=1,2}
{1\over\displaystyle 2\pi (4\pi \kappa^2)^{2\over 3}}
\int_{M^{10}_i}d^{10}x\sqrt g {1\over 4}F_{IJ}^aF^{aIJ} ~.~\,
\end{eqnarray}
In the 11-dimensional 
metric~\footnote{Because we think 11-dimensional
metric is more fundamental than string metric
and Einstein frame, 
 our discussion in this paper use 11-dimensional metric.}, 
the gauge couplings and gravitational
coupling in 4-dimension are~\cite{Witten,TIAN, HLYLZH}:
\begin{eqnarray}
8\pi\,G_{N}^{(4)} &=& {\kappa^2 \over 
{2\pi \rho_p V_p}} ~,~ \, 
\end{eqnarray}
\begin{eqnarray}
\alpha_{\rm GUT} &=&{1\over {2 V_p (1+x)}}\,(4\pi\kappa^2 
)^{2/3} ~,~ \,
\end{eqnarray}
\begin{eqnarray}
\left[\alpha_H \right]_W &=&{1\over {2 V_p (1-x)}}\,(4\pi\kappa^2 
)^{2/3} ~,~\,
\end{eqnarray}
where $x$ is defined by:
\begin{eqnarray}
x &=& \pi^2 {\rho_p \over V_p^{2/3}} 
({\kappa \over 4 \pi })^{2/3} \int_X \omega \wedge 
{{trF \wedge F - {1\over 2} tr R \wedge R}
\over\displaystyle { 8 \pi^2}} ~,~\,
\end{eqnarray}
where $\rho_p$, $V_p$ are the physical eleventh dimension radius
and Calabi-Yau manifold's volume ( which is defined by the middle 
point Calabi-Yau manifold's 
volume between the observable sector and the hidden sector )
respectively.
From above formula, one obtains:
\begin{eqnarray}
x &=& {{\alpha_H \alpha_{GUT}^{-1} - 1} \over\displaystyle
{\alpha_H \alpha_{GUT}^{-1} + 1}} ~.~\,
\end{eqnarray}
The GUT scale $M_{GUT}$ and  the hidden sector GUT scale
$M_H$ when 
the Calabi-Yau manifold is compactified are:
\begin{eqnarray}
M_{\rm GUT}^{-6} &=& V_p ( 1+x) ~,~\,
\end{eqnarray}
\begin{eqnarray}
M_H^{-6} &=& V_p ( 1-x) ~,~\,
\end{eqnarray}
or we can express the $M_H$ as:
\begin{eqnarray}
M_H &=& ({\alpha_H \over\displaystyle 
\alpha_{GUT}})^{1/6} M_{GUT} = ({{1+x} 
\over\displaystyle {1-x}})^{1/6} M_{GUT} ~.~\,
\end{eqnarray}
Noticing that $M_{11} = \kappa^{-2/9}$, we have
\begin{eqnarray}
M_{11} &=& \left[2 (4\pi )^{-2/3}\,  
\, \alpha_{\rm GUT}\right]^{-1/6} M_{GUT}  ~.~\,
\end{eqnarray}
And the physical 
 scale of the eleventh dimension in
the eleven-dimensional metric is:
\begin{eqnarray}
\left[\pi \rho_p\right]^{-1} &=& 
{{8 \pi}\over\displaystyle {1+x}} \left(2 
\alpha_{\rm GUT}\right)^{-3/2}
{{M_{GUT}^3}\over\displaystyle
  {M_{Pl}^2}}~,~\, 
\end{eqnarray}
where $M_{pl}=2.4\times 10^{18}$ GeV.
From the constraints that $M_{GUT}$ and $M_H$ are smaller than
the scale of $M_{11}$, one obtains:
\begin{eqnarray}    
\alpha_{GUT} \leq {{(4 \pi)^{2/3}}\over\displaystyle
2} ~;~
\alpha_H \leq {{(4 \pi)^{2/3}}\over\displaystyle
2} ~,~ \,
\end{eqnarray}
or 
\begin{eqnarray}    
\alpha_{GUT} \leq 2.7 ~;~
\alpha_H \leq 2.7 ~.~ \,
\end{eqnarray}
 For the 
standard embedding,  the upper bound on
x is 0.97 ( $x < 0.97 $ ),
for $\alpha_{GUT} = {1\over 25}$. 

Second,  
let us review the K\"ahler potential, gauge kinetic function, 
 superpotential and soft terms in the simplest compactification
of M-theory on $S^1/Z_2$. 
The K\"ahler potential, gauge kinetic function and  
 superpotential are ~\cite{NOY,LOD}:
\begin{eqnarray}
K &=& \hat K + \tilde K |C|^2 ~,~ \,
\end{eqnarray}
\begin{eqnarray}
\hat K &=&  -\ln\,[S+\bar S]-3\ln\,[T+\bar T] ~,~\,
\end{eqnarray}
\begin{eqnarray}
\tilde K &=& ({3\over\displaystyle {T+\bar T}} +
{\alpha\over\displaystyle {S+\bar S}}) |C|^2  ~,~ \,
\end{eqnarray}
\begin{eqnarray}
Ref^O_{\alpha \beta} &=& Re(S + \alpha T)\, \delta_{\alpha \beta} ~,~\,
\end{eqnarray}
\begin{eqnarray}
Ref^H_{\alpha \beta} &=& Re(S - \alpha T)\, \delta_{\alpha \beta} ~,~\,
\end{eqnarray}
\begin{eqnarray}
W= d_{x y z} C^x C^y C^z ~,~\,
\end{eqnarray}
where $S$, $T$ and $C$ are dilaton, moduli and matter fields 
respectively. $\alpha$ is the next order correction constant 
which is related to the  Calabi-Yau manifold and it is:
\begin{eqnarray}
\alpha={{x ( S + \bar S )} \over\displaystyle { T + \bar T }} 
 ~.~\,
\end{eqnarray}
The non-perturbative superpotential
due to the gaugino condensation is~\cite{LOW, MDRSEW}:
\begin{eqnarray}
W_{np} &=& h~ exp(-{{8 \pi^2}\over\displaystyle C_2(G^H)}
(S- \alpha T)) ~,~\,
\end{eqnarray}
where the group in the hidden sector is $G^H$ and
$C_2(G^H)$ is the
quadratic Casimir of $G^H$. And in this case, $G^H$ is $E_8$
and $C_2(G^H) = 30$. 

With the standard fomulae~\cite{ABIM, VSKJL},
 one can easily obtain the following
soft terms~\cite{JUN,CKM}:
\begin{eqnarray}
M_{1/2}&=&{{\sqrt 3 C_0 M_{3/2}} \over\displaystyle {1+x}}
(\sin\theta e^{-i\theta_S} +{x\over \sqrt 3} \cos\theta
e^{-i\theta_T}) ~,~\,
\end{eqnarray}
\begin{eqnarray}
M_0^2&=& V_0+ M_{3/2}^2 - 
{{3 C_0 M_{3/2}^2} \over\displaystyle 
{(3+x)^2}}(x(6+x) \sin^2\theta +
\nonumber\\&&
 ( 3 + 2x ) \cos^2\theta
- 2 \sqrt 3 x \cos(\theta_S-\theta_T)~\sin\theta ~\cos\theta) ~,~ \,
\end{eqnarray}
\begin{eqnarray}
A&=&- {{\sqrt 3  C_0 M_{3/2}} \over\displaystyle 
{(3+x)}} ((3-2x) \sin\theta e^{-i\theta_S} + \sqrt 3~ x ~\cos\theta
e^{-i\theta_T}) ~,~ \,
\end{eqnarray}
where $M_{3/2}$ is the gravitino
mass, 
\begin{eqnarray}
F^S=\sqrt 3 M_{3/2} C_0 (S+\bar S) \sin\theta e^{-i\theta_S}
~,~\, 
\end{eqnarray}
\begin{eqnarray}
F^T= M_{3/2} C_0 (T+\bar T) \cos\theta e^{-i\theta_T}
~,~\, 
\end{eqnarray}
\begin{eqnarray}
C_0^2= 1+{{V_0} \over {3 M_{3/2}^2}} 
~,~\, 
\end{eqnarray}
and $V_0$ for the
tree level vacuum density.

Generically, the dynamics of the hidden sector may give rise to
both $< F^S >$ and $< F^T > $, but one type of F term often dominates.
 Therefore, we concentrate on the two limiting cases:
dilaton dominant SUSY breaking ($F^T=0$) and the moduli dominant 
SUSY breaking ($F^S=0$)~\cite{ABIM, VSKJL}~\footnote{
In the discussion of the dilaton dominant SUSY breaking
scenario and moduli dominant SUSY breaking scenario,
in order to obtain the simple soft term relations, we set $V_0=0$
, $C_0=1$ and $\theta_S=\theta_T =0$.}:

(I) Dilaton dominant SUSY breaking scenario ($F^T=0$). In this 
case, the soft terms become

\begin{eqnarray}
M_{1/2}&=&{{\sqrt 3  M_{3/2}} \over\displaystyle {1+x}} ~,~\,
\end{eqnarray}
\begin{eqnarray}
M_0^2&=& M_{3/2}^2 - {{ 3  M_{3/2}^2} \over\displaystyle 
{(3+x)^2}} ~x ~(6+x) ~,~  \,
\end{eqnarray}
\begin{eqnarray}
A&=&- {{\sqrt 3  M_{3/2}} \over\displaystyle 
{3+x }} (3-2x) ~.~\,
\end{eqnarray}

(II) Moduli dominant SUSY breaking scenario ( $F^S = 0$ ). In 
this case, the soft terms become

\begin{eqnarray}
M_{1/2}&=&{x\over\displaystyle {1+x}} M_{3/2} ~,~ \,
\end{eqnarray}
\begin{eqnarray}
M_0&=& {x\over\displaystyle {3+x}} M_{3/2} ~,~\,
\end{eqnarray}
\begin{eqnarray}
A&=&- {{ 3x } \over\displaystyle {3+x}} M_{3/2} ~.~\,
\end{eqnarray}
Therefore, we have
\begin{eqnarray}
 M_0/A=-1/3 ~;~ 3 \geq M_{1/2}/M_0 \geq 2 ~.~\,
\end{eqnarray} 

However, the previous calculation is expanded in x
( small x is considered) and
the Lagrangian, K\"ahler potential and gauge kinetic function
are at order of x. Therefore, to be consistent, we should 
keep the scale, gauge couplings and
 soft terms at order of x, because there may exist other
high order correction ( $x^2$ and higher ) to 
the Lagrangian, or K\"ahler potential and gauge kinetic function.
Considering $x^n = 0$ for n $>$ 1, the following previous scale and gauge 
coupling equations
 need to be changed:
\begin{eqnarray}
\alpha_{\rm GUT} &=&{{1-x} \over\displaystyle {2 V_p }}\,(4\pi\kappa^2 
)^{2/3} ~,~ \,
\end{eqnarray}
\begin{eqnarray}
\left[\alpha_H \right]_W &=&{{1+x} \over\displaystyle {2 V_p }}\,(4\pi\kappa^2 
)^{2/3} ~,~\,
\end{eqnarray}
\begin{eqnarray}
M_H=(1+x/3) M_{GUT}
~,~\,
\end{eqnarray}
\begin{eqnarray}
\left[\pi \rho_p\right]^{-1} &=& 
8 \pi (1-x) \left(2 
\alpha_{\rm GUT}\right)^{-3/2}
{{M_{GUT}^3}\over\displaystyle
  {M_{Pl}^2}}~,~\, 
\end{eqnarray}
and the next order soft  terms are:
\begin{eqnarray}
M_{1/2}&=&{\sqrt 3 C_0  M_{3/2}} 
( (1-x) \sin\theta e^{-i\theta_S} +{x\over \sqrt 3} \cos\theta
e^{-i\theta_T} ) ~,~\,
\end{eqnarray}
\begin{eqnarray}
M_0^2&=& V_0 + M_{3/2}^2 - 
{{C_0^2  M_{3/2}^2} \over\displaystyle 
3}(6x \sin^2\theta +
\nonumber\\&&
  3  \cos^2\theta
- 2 \sqrt 3 x \cos(\theta_S -\theta_T)
 ~\sin\theta ~\cos\theta) ~,~ \,
\end{eqnarray}
\begin{eqnarray}
A&=&- {\sqrt 3 C_0  M_{3/2}} 
( (1-x) \sin\theta e^{-i\theta_S} +{x\over \sqrt 3} \cos\theta
e^{-i\theta_T} )
 ~.~ \,
\end{eqnarray}
Obviously, we have $M_{1/2} = -A$. And we obtain the soft terms
in the two limiting case:

(I) Dilaton dominant SUSY breaking scenario ($F^T=0$): 
\begin{eqnarray}
M_{1/2}&=&\sqrt 3  M_{3/2} (1-x)  ~,~\,
\end{eqnarray}
\begin{eqnarray}
M_0&=& (1-x) M_{3/2} ~,~  \,
\end{eqnarray}
\begin{eqnarray}
A&=&- \sqrt 3  M_{3/2} (1-x) ~.~\,
\end{eqnarray}
Soft term relations are the same as those obtained in
the weakly coupled string
with dilaton dominant SUSY breaking: $M_{1/2}=-A=\sqrt 3 M_0$.

(II) Moduli dominant SUSY breaking scenario ( $F^S = 0$ ):

\begin{eqnarray}
M_{1/2}&=& x  M_{3/2} ~,~ \,
\end{eqnarray}
\begin{eqnarray}
M_0 = 0 ~,~\,
\end{eqnarray}
\begin{eqnarray}
A&=& -x M_{3/2} ~.~\,
\end{eqnarray}
Soft term relations: $M_{1/2}=-A, M_0 =0$ are similar to the
no-scale case~\cite{no-scale}: 
$M_{1/2} \neq 0$ and $M_0=A=0$, for A is not very important 
in the RGE running.

Because in general, we have $M_{1/2} =-A$. We might need to
know the relation between the magnitude of $M_{1/2}$ and
that of $M_0$. Taking $V_0=0, C_0=1$, 
we can express $|M_{1/2}|^2$ as following:
\begin{eqnarray}
|M_{1/2}|^2= 3 M_0^2 + M_{2/3}^2 x^2 \left(3 \sin^2\theta
+ \cos^2\theta 
- 2 \sqrt 3 \cos(\theta_S -\theta_T)
 ~\sin\theta ~\cos\theta \right)
~.~\,
\end{eqnarray}
We obtain that 
\begin{eqnarray}
M_{1/2} = -A ~,~
|{{M_{0}}\over\displaystyle {M_{1/2}}}| \leq 
{1\over\displaystyle \sqrt 3}
 ~,~\,
\end{eqnarray} 
for the last term in eq. (51) is obvious positive,
and only when $\theta_S-\theta_T=0$ ( $\pi $) and
$\theta = {\pi \over\displaystyle 6}$
or $\theta = {{ 7 \pi} \over\displaystyle 6}$
( $\theta = {{5 \pi} \over\displaystyle 6}$
or $\theta = {{11 \pi} \over\displaystyle 6}$), 
$|M_{1/2}| = \sqrt 3 |M_0|$.
In fig. 1, 
choosing $\theta_S=\theta_T=0$,
we draw ${{M_{0}}\over\displaystyle {M_{1/2}}}$
versus $\theta$ by taking x =0.05, 0.1, 0.2, 0.3 which are
represented by the solid, dots, dotdash, and dashes lines, respectively.
We can see that the deviation from $\pm {1 \over\displaystyle \sqrt 3}$
is large when $\theta$ closes to 0 or $\pi$, or $x$ is large.  
The soft terms  are different from those obtained in the
weakly coupled string where one has~\cite{ABIM}:
\begin{eqnarray}
M_{1/2} = -A ~,~
|{{M_{0}}\over\displaystyle {M_{1/2}}}| =
{1\over\displaystyle \sqrt 3}
 ~.~\,
\end{eqnarray}

\subsection{ Toy Model Compactification  and its Phenomenology}

If we did not consider the higher order
correction to the bulk fields, one can easily 
write down the most general parametrized K\"ahler potential,
gauge kinetic function and non-perturbative superpotential. 
For the simplest compactification, they are~\cite{JUNJUN}:
\begin{eqnarray}
K &=& \hat K + \tilde K |C|^2 ~,~ \,
\end{eqnarray}
\begin{eqnarray}
\hat K &=&  -\ln\,[S+\bar S]-3\ln\,[T+\bar T] ~,~\,
\end{eqnarray}
\begin{eqnarray}
\tilde K &=& \left(1+\sum_{i=1}^{\infty} c_i 
\left({{\alpha (T+\bar T)} \over\displaystyle {S+\bar S}} \right)^i
\right)
({3\over\displaystyle {T+\bar T}})
 |C|^2  ~,~ \,
\end{eqnarray}
\begin{eqnarray}
f^O_{\alpha \beta} &=& S
\left(1+\sum_{i=1}^{\infty} d_i 
\left({{\alpha T} \over\displaystyle {S}} \right)^i
\right)\, \delta_{\alpha \beta} ~,~\,
\end{eqnarray}
\begin{eqnarray}
f^H_{\alpha \beta} &=& S
\left(1+\sum_{i=1}^{\infty} d_i 
\left({{-\alpha T} \over\displaystyle S} \right)^i
\right)\, \delta_{\alpha \beta} ~,~\,
\end{eqnarray}
\begin{eqnarray}
W_{np} &=& h~ exp\left(-{{8 \pi^2}\over\displaystyle C_2(G^H)}
S \left(1+\sum_{i=1}^{\infty} d_i 
\left({{-\alpha T} \over\displaystyle S} \right)^i
\right)\right) ~.~\,
\end{eqnarray} 
Using standard method~\cite{ABIM, VSKJL},
 one can easily calculate the soft terms. 
 But, here, one introduce many parameters, which is useless to
 the low energy phenomenology analysis. Therefore, we 
 construct a toy simple model as an example which contains
 high order correction. 

In fact, in order to consider high order correction to the 
scale, gauge couplings and soft terms in detail, we need to consider
the higher order Lagrangian. However, the original 
Lagrangian is at order of $\kappa^{2/3}$
~\cite{HW}, and the Witten's solution to
the compactification of M-theory on $S^1/Z_2$ which have N=1 supersymmetry in
4-dimension just considered the next order expansion of the
metric~\cite{Witten}. Therefore, it is really very tough to consider the
higher order terms realistically. The ansatz of our toy model is that,
we just consider the Lagrangian which was obtained by Horava and
Witten, we do not consider the higher order expansion of 
the metric and higher order correction to the
bulk fields. Under this ansatz, we can discuss the 
higher order correction to the scale, gauge couplings,
 K\"ahler potential,  gauge 
kinetic function and superpotential.

The bosonic part of the
eleven-dimensional supergravity Lagrangian is given by~\cite{HW}
\begin{eqnarray}
L_B&=&{1\over \kappa^2}\int_{M^{11}}d^{11}x\sqrt g
\left(-{1\over 2}R -{1\over 48}G_{IJKL}G^{IJKL}
\right.\left. -{\sqrt 2\over 3456}
\epsilon^{I_1I_2\dots I_{11}}C_{I_1I_2I_3}G_{I_4\dots I_7}G_{I_8\dots
I_{11}}\right)\nonumber\\
&&\qquad\qquad-\sum_{i=1,2}
{1\over\displaystyle 2\pi (4\pi \kappa^2)^{2\over 3}}
\int_{M^{10}_i}d^{10}x\sqrt g {1\over 4}F_{IJ}^aF^{aIJ}
~,~\,
\end{eqnarray}
where $ G_{11\,IJK}=\left(\partial_{11}C_{IJK}\pm 23 \,\,{\rm
permutations}\right)
+{\kappa^2\over \sqrt 2 \lambda^2}\delta(x^{11})\omega_{IJK}$,
$\lambda^2=2\pi (4 \pi \kappa^2)^{2/3}$.

In this paper, we consider the compactification on the
Calabi-Yau manifold with
Hodge-Betti numbers $h_{(1,1)}=1$ and $h_{(2,1)}=0$.
For the zeroth-order metric, there are only
one dilaton and one
 modulus controlling the overall size of the Calabi-Yau space and the
length of the orbifold interval. We write 
\begin{eqnarray}
  ds^2 = g_{\mu \nu}^{(0)} dx^\mu dx^\nu 
       + e^{2a} g_{AB}^{(0)} dx^Adx^B + e^{2c} (dx^{11})^2 ~,~\,
\end{eqnarray}
so that the physical Calabi-Yau volume $V_p$ is $e^{6a}V$ and the 
physical length of the orbifold
interval $\pi \rho_p$ is $e^c \pi \rho$.
For the next order correction, because the massive modes
are supressed by the factor ${V_p^{1/6} \over\displaystyle {\pi \rho_p}}$
which is at about 0.1 order and very small when we consider the 
intermediate unification,
we just consider the massless mode. And the deformed
metric are~\cite{LOD}:
\begin{eqnarray}
  ds^2 &=& (1+\beta e^{c-4a} (|{{2 x^{11}}\over {\pi \rho}}|-1))
    g_{\mu \nu}^{(0)} dx^\mu dx^\nu 
       + (1-\beta e^{c-4a} (|{{2 x^{11}}\over {\pi \rho}}|-1))
\nonumber\\&&       
       e^{2a} g_{AB}^{(0)} dx^Adx^B + 
      (1 - 2 \beta e^{c-4a} (|{{2 x^{11}}\over {\pi \rho}}|-1)) 
       e^{2c} (dx^{11})^2 ~,~\,
\end{eqnarray}
where

\begin{eqnarray}
\beta &=& {1\over 3} \pi^2 {\rho \over V^{2/3}} 
({\kappa \over 4 \pi })^{2/3} \int_X \omega \wedge 
{{trF \wedge F - {1\over 2} tr R \wedge R}
\over\displaystyle { 8 \pi^2}} ~.~\,
\end{eqnarray}
The M-theory limit is $\beta e^{c-4a} = \pm {1\over 2}$.
In order to keep $g_{11, 11}$ positive,
we require that $ - {1\over 2} <\beta e^{c-4a} < {1\over 2} $, 
which is enough to keep  
the physical Calabi-Yau manifold's volume non-zero at
any point along the eleventh dimension.

Defining  
\begin{eqnarray}
 C_{AB11} = {1\over 6} \chi \omega_{AB}
~,~
 C_{\mu \nu 11} = {1\over 6} B_{\mu\nu} 
 ~,~\,
\end{eqnarray}
and
\begin{eqnarray}
\partial_{[ \mu} B_{\nu \rho ]}=
{1\over 3} e^{-12 a} {\epsilon_{\mu \nu \rho}}^{\delta}
\partial_{\delta} \sigma ~,~
\hat c = c +2 a
 ~,~\,
\end{eqnarray}
one can obtain the Lagrangian in 4-dimension to the zeroth
order~\cite{TIAN, LOD}:
\begin{eqnarray}
S^1_0 &=& {{\pi \rho V}\over\displaystyle \kappa^2}
\int_{M^4}\sqrt {-g}\left[
-R-18\partial_\mu a \partial^\mu a -
{3\over 2} \partial_\mu \hat c \partial^\mu \hat c  -
3e^{-2 \hat c }\partial_\mu \chi \partial^\mu \chi
 \right.\nonumber\\&&\left.
-e^{-12a}\partial_\mu \sigma \partial^\mu \sigma
\right]
 ~,~\,
\end{eqnarray}
\begin{eqnarray}
S^2_0 &=& 
{{\pi \rho V}\over\displaystyle {\kappa^2}}
\int_{M^4}\sqrt {-g}\left[ -3e^{-\hat c}
            D_\mu C D^\mu \bar C -{{3 i}\over {\sqrt{2}}}
            e^{-2\hat{c}}
           \left(\bar C D_\mu C -CD_\mu \bar C \right)
           \partial^\mu\chi 
 \right.\nonumber\\&&\left.           
-{{3k^2}\over 4} e^{-2\hat c -6a}
              |d_{pqr}C^pC^q|^2-
{{3k^2}\over {32}}e^{-2\hat c-6a}
              (\bar{C}T^iC)^2
 \right.\nonumber\\&&\left.
-{1\over 4} e^{6 a} F^O_{\mu \nu} F^{O \mu \nu} 
-{1\over 4} e^{6 a} F^H_{\mu \nu} F^{H \mu \nu} 
-{{\sqrt 2 \sigma }\over 4}
F^O_{\mu \nu} \widetilde {F^{O \mu \nu}}
-{{\sqrt 2 \sigma }\over 4}
F^H_{\mu \nu} \widetilde {F^{H \mu \nu}}
\right]  ~,~\,
\end{eqnarray}
\begin{eqnarray}
S^3_0&=&
{{\pi \rho V}\over\displaystyle {\kappa}^2}
\int_{M^4}\sqrt{-g}\left[
             {{3}\over {8}}
             e^{-2 \hat c}(C^2 D_\mu \bar C D^{\mu} \bar C
             + {\bar C} ^2
             D_\mu C D^\mu C -2|C|^2 D_\mu C D^\mu \bar C ) 
 \right.\nonumber\\&&\left.             
 -{{3 k^2}\over 8}e^{-3\hat c -6a}
|d_{pqr}C^pC^qC^r|^2\right]
 ~,~\,
\end{eqnarray}
where $k=4\sqrt{2 \rho }\pi (4\pi /\kappa )^{1/3}$
and for the gauge fields and gauge matter fileds, we rescale
them as following~\cite{TIAN, LOD}:
\begin{eqnarray}
 C^p \rightarrow \pi\sqrt{2\rho}\left({{4\pi}\over {\kappa}}\right)^{1/3} C^p
 ~,~
 A_{\mu} \rightarrow \pi\sqrt{2\rho}\left({{4\pi}\over {\kappa}}\right)^{1/3} 
 A_{\mu} ~.~\,
\end{eqnarray}
In above equations, $S^1_0$ is from the bulk Lagrangian,
$S^2_0$ is from the boundary Lagrangian and  
$S^3_0$ is from the boundary Lagrangian with additional 
$\delta (0)$ factor which is not well defined.
Considering the higher order correction from the deformed metric to the
boundary terms, we obtain:
\begin{eqnarray}
S^2 &=& 
{{\pi \rho V}\over\displaystyle {\kappa^2}}
\int_{M^4}\sqrt {-g}\left[ -3e^{-\hat c} \Lambda_+^2 \Lambda_-
            D_\mu C D^\mu \bar C -
            {{3 i}\over {\sqrt{2}}} \Lambda_+ \Lambda_-
            e^{-2\hat{c}}
           \left(\bar C D_\mu C -CD_\mu \bar C \right)
           \partial^\mu\chi 
 \right.\nonumber\\&&\left.           
-{{3k^2}\over 4} \Lambda_+ \Lambda_-^2 e^{-2\hat c -6a}
              |d_{pqr}C^pC^q|^2-
{{3k^2}\over {32}} \Lambda_+ \Lambda_-^2 e^{-2\hat c-6a}
              (\bar{C}T^iC)^2
 \right.\nonumber\\&&\left.
-{1\over 4}\Lambda_+^3 e^{6 a} F^O_{\mu \nu} F^{O \mu \nu} 
-{1\over 4} \Lambda_-^3 e^{6 a} F^H_{\mu \nu} F^{H \mu \nu} 
-{{\sqrt 2 \sigma }\over 4} \Lambda_+^3
F^O_{\mu \nu} \widetilde {F^{O \mu \nu}}
 \right.\nonumber\\&&\left.
-{{\sqrt 2 \sigma }\over 4} \Lambda_-^3
F^H_{\mu \nu} \widetilde {F^{ H \mu \nu}}
\right]  ~,~\,
\end{eqnarray} 
\begin{eqnarray}
S^3&=&
{{\pi \rho V}\over\displaystyle {\kappa}^2}
\int_{M^4}\sqrt{-g}\left[
             {{3}\over {8}} \Lambda_+ \Lambda_-
             e^{-2 \hat c}(C^2 D_\mu \bar C D^{\mu} \bar C
             + {\bar C} ^2
             D_\mu C D^\mu C -2|C|^2 D_\mu C D^\mu \bar C ) 
 \right.\nonumber\\&&\left.             
 -{{3 k^2}\over 8} \Lambda_-^2
 e^{-3\hat c -6a}
|d_{pqr}C^pC^qC^r|^2\right] {1\over {(1+ 2 \beta e^{c-4a})^{1/2}}}
 ~,~\,
\end{eqnarray}
where
\begin{eqnarray}
\Lambda_+ = 1 + \beta e^{c-4a} ~,~
\Lambda_- = 1 - \beta e^{c-4a} ~.~\,
\end{eqnarray}
Because $S^3$ is proportional to $\delta (0)$ in the
original Lagrangian and its order is $\kappa^{4/3}$, 
we use $S^2$ to obtain the
 K\"ahler potential, gauge kinetic function and  
the superpotential in this toy compactification:
\begin{eqnarray}
K &=& \hat K + \tilde K |C|^2 ~,~ \,
\end{eqnarray}
\begin{eqnarray}
\hat K &=&  -\ln\,[S+\bar S]-3\ln\,[T+\bar T] ~,~\,
\end{eqnarray}
\begin{eqnarray}
\tilde K &=& {3\over\displaystyle {T+\bar T}}
\left(1+{{\beta (T+\bar T)}\over\displaystyle 
{S+\bar S}}\right)^2
\left(1-{{\beta (T+\bar T)}\over\displaystyle 
{S+\bar S}}\right)
 ~,~ \,
\end{eqnarray}
\begin{eqnarray}
f^O_{\alpha \beta} &=& S \left(1+{{\beta T}\over\displaystyle 
{S}}\right)^3
 \delta_{\alpha \beta} ~,~\,
\end{eqnarray}
\begin{eqnarray}
f^H_{\alpha \beta} &=& S \left(1-{{\beta T}\over\displaystyle 
{S}}\right)^3
 \delta_{\alpha \beta} ~,~\,
\end{eqnarray}
\begin{eqnarray}
W= \left(1+{{\beta T}\over\displaystyle 
{S}}\right)^{3/2}
\left(1-{{\beta T}\over\displaystyle 
{S}}\right)^{3/2} k d_{x y z} C^x C^y C^z ~,~\,
\end{eqnarray}
\begin{eqnarray}
W_{np} &=& h~ exp\left(-{{8 \pi^2}\over\displaystyle C_2(G^H)}
S \left(1-{{\beta T}\over\displaystyle 
{S}}\right)^3
 \right) ~.~\,
\end{eqnarray} 
where
\begin{eqnarray}
 S=e^{6a}+i\sqrt 2 \sigma ~,~ T = e^{\hat c}+i\sqrt 2\chi
~.~\,
\end{eqnarray} 
One can easily prove that, when $\beta$ is very small, i. e., 
 taking $\beta^2 =$0, one obtains the 
same results as those obtained previously~\cite{NOY, LOD}.

Now, we will consider the phenomenology in the toy compactification.
First, let us discuss the scale and gauge couplings,
\begin{eqnarray}
8\pi\,G_{N}^{(4)} &=& {\kappa^2 \over 
{2\pi \rho_p V_p \delta}} ~,~ \, 
\end{eqnarray}
\begin{eqnarray}
\alpha_{\rm GUT} &=&{1\over {2 V_p (1+y)^3}}\,(4\pi\kappa^2 
)^{2/3} ~,~ \,
\end{eqnarray}
\begin{eqnarray}
\alpha_H &=&{1\over {2 V_p (1-y)^3}}\,(4\pi\kappa^2 
)^{2/3} ~,~\,
\end{eqnarray}
where 
\begin{eqnarray}
y&=& {{\beta (T+\bar T)} \over\displaystyle {S+\bar S}} =
{{\alpha_H^{1/3} \alpha_{GUT}^{-1/3} - 1} \over\displaystyle
{\alpha_H^{1/3} \alpha_{GUT}^{-1/3} + 1}}
~,~\,
\end{eqnarray} 
\begin{eqnarray}
\delta &=& {{\int_{-1}^{1} \sqrt {1- 2 y x} (1+yx)^2 (1-yx)^3 dx}
\over\displaystyle {\int_{-1}^{1} \sqrt {1-2 y x} dx }}
~.~\,
\end{eqnarray} 
By the way, comparing the compactification in last 
subsection, we have: $y={x \over\displaystyle 3}, 
\beta={\alpha \over\displaystyle 3}$.

The GUT scale $M_{GUT}$ and  the hidden sector GUT scale
$M_H$ when 
the Calabi-Yau manifold is compactified are:
\begin{eqnarray}
M_{\rm GUT}^{-6} &=& V_p ( 1+y)^3 ~,~\,
\end{eqnarray}
\begin{eqnarray}
M_H^{-6} &=& V_p ( 1-y)^3 ~,~\,
\end{eqnarray}
or we can express the $M_H$ as:
\begin{eqnarray}
M_H  = ({{1+y} 
\over\displaystyle {1-y}})^{1/2} M_{GUT} ~,~\,
\end{eqnarray}
\begin{eqnarray}
M_{11} &=& \left[2 (4\pi )^{-2/3}\,  
\, \alpha_{\rm GUT}\right]^{-1/6} M_{GUT}  ~.~\,
\end{eqnarray}
And the physical 
 scale of the eleventh dimension is:
\begin{eqnarray}
\left[\pi \rho_p\right]^{-1} &=& 
{{8 \pi \delta }\over\displaystyle {(1+y)^3}} \left(2 
\alpha_{\rm GUT}\right)^{-3/2}
{{M_{GUT}^3}\over\displaystyle
  {M_{Pl}^2}}~.~\, 
\end{eqnarray}

In order to keep $g_{11, 11}$ positive, we require that
$  -{1\over 2} < y < {1\over 2}$. In addition, in fig. 2,
we plot the $\delta$ versus $y$, and we find that $\delta$
is about 1. So, $\delta$ will not  change previous scale
picture~\cite{JUNJUN, HLYLZH, DGCMUNOZ}.
And the bounds on $M_H$ and $\alpha_H$ are:
\begin{eqnarray}
{1 \over {27}} \alpha_{GUT} < \alpha_H < 27 \alpha_{GUT}
~,~\, 
\end{eqnarray}
\begin{eqnarray}
{1\over \sqrt 3} M_{GUT} < M_H < \sqrt 3 M_{GUT}
~.~\, 
\end{eqnarray}
In addition, because $  -{1\over 2} < y < {1\over 2}$,
the discussions of the
scale and gauge couplings are similar to those
in ~\cite{JUNJUN, HLYLZH, DGCMUNOZ}
 with small $x$. So, we will not redo the discussion
here again.

Second, let us discuss the soft terms from above K\"ahler  
potential and gauge kinetic function.
Using standard tree level fomulae
\cite{ABIM, VSKJL}, we obtain the following
soft terms:
\begin{eqnarray}
M_{1/2} &=& {{\sqrt 3 C_0 M_{3/2}} \over\displaystyle {1+y}}\left[
\sin\theta (1-2 y) e^{-i\theta_S}
+ \sqrt 3 y \cos\theta e^{-i \theta_T } \right]
~,~\, 
\end{eqnarray}
\begin{eqnarray}
M_0^2&=&M_{3/2}^2 +V_0 -{{M_{3/2}^2 C_0^2}
\over\displaystyle {(1-y^2)^2}}
\nonumber\\&&    
\left( 3 y (2-9 y +3 y^3) \sin^2\theta 
+(1-5 y^2+2 y^3-2 y^4) \cos^2\theta
\right.\nonumber\\&&\left.
+2 \sqrt 3 y \sin\theta \cos\theta 
\cos(\theta_S-\theta_T) (-1 + 6 y - y^2)
\right)
~,~\, 
\end{eqnarray}
\begin{eqnarray}
A&=& -{{\sqrt 3} \over\displaystyle {1-y^2}}
M_{3/2} C_0 (1-3 y + 8 y^2) \sin\theta e^{-i\theta_S}
\nonumber\\&&
-{{3 y} \over\displaystyle {1-y^2}}
M_{3/2} C_0 (1-3 y) \cos\theta e^{-i\theta_T}
~,~\, 
\end{eqnarray}
where 
\begin{eqnarray}
F^S=\sqrt 3 M_{3/2} C_0 (S+\bar S) \sin\theta e^{-i\theta_S}
~,~\, 
\end{eqnarray}
\begin{eqnarray}
F^T= M_{3/2} C_0 (T+\bar T) \cos\theta e^{-i\theta_T}
~,~\, 
\end{eqnarray}
\begin{eqnarray}
C_0^2= 1+{{V_0} \over {3 M_{3/2}^2}}
~,~\, 
\end{eqnarray}
 and $V_0$ for the
tree level vacuum density.

Now, we discuss the numerical results for the soft terms
by taking $V_0=0, C_0=1$, and $\theta_S=\theta_T=0$.
First, we compare the soft terms. We denote the
original soft terms ( eq.s (28-30) ) as scenario O, the soft terms
( eq.s (45-47) ) at next to the leading order
which is simple as scenario S, the soft terms
( eq.s (96-98))  in above toy compactification
as scenario T. Choosing $\theta = {\pi \over 4}$, we draw
the soft terms: $M_{1/2}, M_0, A$ versus $x$ in fig. 3, fig. 4,
and fig. 5 respectively.
We notice that, when $x < 0.2 $, three scenarios agree very well,
when $x > 0.5$, we can see the obvious differences among them.
In addition, we can compare the magnitudes of the soft terms in
these three scenarios:
$M_{1/2}^S < M_{1/2}^T < M_{1/2}^O$,
$M_0^S < M_0^O < M_0^T$,
$|A|^S < |A|^O < |A|^T$.
Therefore, if we discuss M-theory phenomenology from the
soft terms, we should keep in the small $x$ region
in order to be consistent, because we do 
not know the higher order correction, although in realistic model, 
$x$ might be large.     
Second, we analyze the soft terms  in the toy compactification,
we drow the soft terms versus $\theta$ by choosing $y=0.15,
0.2, 0.3, 0.4, 0.5$ in fig. 6, 7,  8, 9, 10, respectively.
We notice that, if y is small ( $ y < 0.2$ ), in large
parameter space, 
the magnitude of the gaugino mass
is larger than that of the scalar mass, if y is large ( $ y > 0.3$ ), 
in the most of the parameter space, the magnitude of the gaugino mass
is smaller than that of the scalar mass. However, in previous 
soft term analysis in standard 
embedding, the magnitude of the gaugino mass is often larger
than that of the scalar mass~\cite{CKM}. 
We also discuss the case that $y < 0$, where we just
use $y=-0.3$ as an example in fig. 11. 

\section{Scale, Gauge Couplings,
Soft Terms and Toy Compactification in Non-Standard Embedding}

\subsection{ Scale, Gauge Couplings,
K\"ahler Function and Soft Terms Revisit}
The non-standard embedding without 5-branes is similar to the
standard embedding in section 2.1.
In general, non-standard embedding includes 5-brane correction,
which introduces additional moduli fields $Z_n$, whose real parts
are the 
5-brane positions along the eleventh dimension.
The K\"ahler potential, gauge kinetic function and superpotential
are~\cite{NSLOW}:
\begin{eqnarray}
K &=& \hat K + \tilde K |C|^2 ~,~ \,
\end{eqnarray}
\begin{eqnarray}
\hat K &=&  -\ln\,[S+\bar S]-3\ln\,[T+\bar T] +K_5 ~,~\,
\end{eqnarray}
\begin{eqnarray}
\tilde K &=& ({3\over\displaystyle {T+\bar T}} +
{3 \epsilon \zeta \over\displaystyle {S+\bar S}}) |C|^2  ~,~ \,
\end{eqnarray}
\begin{eqnarray}
f^O_{\alpha \beta} &=& \left(S + 3 \epsilon T
(\beta^0+ \sum_{n=1}^{N} (1-Z_n)^2 \beta^n)
\right) \, \delta_{\alpha \beta} ~,~\,
\end{eqnarray}
\begin{eqnarray}
f^H_{\alpha \beta} &=& \left( S + 3 \epsilon T
(\beta^{N+1} + \sum_{n=1}^{N} (Z_n)^2 \beta^n )\right)
\, \delta_{\alpha \beta} ~,~\,
\end{eqnarray}
\begin{eqnarray}
W= d_{x y z} C^x C^y C^z ~,~\,
\end{eqnarray}
\begin{eqnarray}
W_{np} &=& h~ exp\left(-{{8 \pi^2}\over\displaystyle C_2(G^H)}
 (S + 3 \epsilon T
(\beta^{N+1} + \sum_{n=1}^{N} (Z_n)^2 \beta^n ))\right) ~,~\,
\end{eqnarray} 
where
\begin{eqnarray}
\epsilon=({\kappa \over {4 \pi}})^{2/3} {{2 \pi^2 \rho}
\over\displaystyle V^{2/3}}~,~\,
\end{eqnarray}
\begin{eqnarray}
\zeta=\beta^0 + \sum_{n=1}^N \left(1- {1\over 2} (Z_n + {\bar Z_n})
\right)^2 \beta^n
~,~\,
\end{eqnarray}
and $K_5$ is the K\"ahler potential for the moduli fields $Z_n$.
In addition,   
$\beta^0, \beta^{N+1}$ are the instanton number in the observable
sector and hidden sector, $\beta^n$ where n=1, ..., N is the 
magnetic charge on each 5-brane. The cohomology constrants on $\beta^i$
is:
\begin{eqnarray} 
\sum_{n=0}^{N+1} \beta^i = 0 
~,~\,
\end{eqnarray}
which means the net charge should be zero.

With assumption: $< Z_n > = < ReZ_n >$, $< S > = < \bar S >$,
$< T > = < \bar T >$, we discuss the
scale and gauge couplings first~\cite{DGCMUNOZ}:
\begin{eqnarray}
8\pi\,G_{N}^{(4)} &=& {\kappa^2 \over 
{2\pi \rho_p V_p \delta_{5b}}} ~,~ \, 
\end{eqnarray}
\begin{eqnarray}
\alpha_{\rm GUT} &=&{1\over {2 V_p (1+e)}}\,(4\pi\kappa^2 
)^{2/3} ~,~ \,
\end{eqnarray}
\begin{eqnarray}
\alpha_H &=&{1\over {2 V_p (1+e_h)}}\,(4\pi\kappa^2 
)^{2/3} ~,~\,
\end{eqnarray}
where 
\begin{eqnarray}
\delta_{5b} &=& {{\int_{0}^{\pi \rho}   (1-2 y_{5b} ) dx^{11}}
\over\displaystyle {\int_{0}^{\pi \rho} (1-y_{5b} ) dx^{11} }}
~.~\,
\end{eqnarray}
\begin{eqnarray} 
 e = {{3 \epsilon \zeta (T + \bar T)}\over\displaystyle
 {S+ \bar S}} 
~,~\,
\end{eqnarray}
\begin{eqnarray} 
 e_h = {{3 \epsilon  (T + \bar T)} \over\displaystyle
 {S+ \bar S}} 
(\beta^{N+1} + \sum_{n=1}^{N} ({1\over 2} (Z_n+\bar Z_n))^2 \beta^n )
~,~\,
\end{eqnarray} 
and where
\begin{eqnarray}
y_{5b}&=& {{\epsilon (T+\bar T)}\over\displaystyle {S+\bar S}}
\left[ 2 \sum_{m=0}^n \beta^m (|ReZ|-
ReZ_m)-\sum_{m=0}^{N+1} ( (ReZ_m)^2 - 2ReZ_m ) \beta^m
\right] 
~,~\,
\end{eqnarray} 
in the interval $ReZ_n \leq |ReZ| \leq ReZ_{n+1}$. And
the $ReZ, ReZ_n$ are defined as:
\begin{eqnarray}
Re Z = {{x^{11}}\over\displaystyle {\pi \rho}}~,~
ReZ_n=  {{x^{11}_n}\over\displaystyle {\pi \rho}}
~,~\,
\end{eqnarray}
where n= 1, ..., N. 
$y_{5b}$ is a function of above $x$, and in
general,  $\delta_{5b}$ will not be 1 in non-standard embedding with
five branes. However, $\delta_{5b}$ may close to 1 for we
consider the small next order correction, so, it will not
change the previous scale picture~\cite{JUNJUN, HLYLZH, DGCMUNOZ}. 
 
The GUT scale $M_{GUT}$ and  the hidden sector GUT scale
$M_H$ when 
the Calabi-Yau manifold is compactified are:
\begin{eqnarray}
M_{\rm GUT}^{-6} &=& V_p (1+e) ~,~\,
\end{eqnarray}
\begin{eqnarray}
M_H^{-6} &=& V_p (1+e_h) ~,~\,
\end{eqnarray}
or we can express the $M_H$ as:
\begin{eqnarray}
M_H  = ({{1+e} 
\over\displaystyle {1+e_h}})^{1/6} M_{GUT} ~,~\,
\end{eqnarray}
\begin{eqnarray}
M_{11} &=& \left[2 (4\pi )^{-2/3}\,  
\, \alpha_{\rm GUT}\right]^{-1/6} M_{GUT}  ~.~\,
\end{eqnarray}
And the physical 
 scale of the eleventh dimension is:
\begin{eqnarray}
\left[\pi \rho_p\right]^{-1} &=& 
{{8 \pi \delta_{5b} }\over\displaystyle {1+e}} \left(2 
\alpha_{\rm GUT}\right)^{-3/2}
{{M_{GUT}^3}\over\displaystyle
  {M_{Pl}^2}}~.~\, 
\end{eqnarray}
Second, we can also obtain the following soft terms~\cite{TKJKHS,
DGCMUNOZ}:
\begin{eqnarray} 
M_{1/2} = {1\over {1+e}} ({\tilde F}^S + e {\tilde F}^T
+ e {\tilde F}^n \zeta_n )
~,~\,
\end{eqnarray}
\begin{eqnarray}
M_0^2 &=& V_0 + M_{3/2}^2 -
{1\over {(3+e)^2}} \left(
e (6+e) |{\tilde F}^S|^2 
 \right.\nonumber\\&&\left.  
+ 3 ( 3 + 2 e ) |{\tilde F}^T|^2 
- 6e Re{\tilde F}^S {\bar {\tilde F}}^{\bar T}
\right.\nonumber\\&&\left.  
+(e \zeta (3+e) \zeta_{n \bar m} -
e^2 \zeta_n \zeta_{\bar m}) {\tilde F}^n {\bar {\tilde F}}^{\bar m}
\right.\nonumber\\&&\left.  
-6 e \zeta_{\bar n} Re{\tilde F}^S {\bar {\tilde F}}^{\bar n}  
+6 e \zeta_{\bar n} Re{\tilde F}^T {\bar {\tilde F}}^{\bar n} 
\right) 
~,~\,
\end{eqnarray}
\begin{eqnarray}
A&=& -{1\over {3+e}} \left( (3-2e) {\tilde F}^S
+ 3 e {\tilde F}^T
\right.\nonumber\\&&\left. 
+(3 e \zeta_n - (3+e) \zeta K_{5, n} )   
 {\tilde F}^n \right)
~,~\,
\end{eqnarray}
where
\begin{eqnarray} 
V_0= |{ F^S \over\displaystyle {S+\bar S}}|^2 + 
3  |{ F^T \over\displaystyle {T+\bar T}}|^2 + 
\bar F^{\bar n} F^m K_{5, \bar n m} - 3 M_{3/2}^2
~,~\,
\end{eqnarray}
and
\begin{eqnarray} 
  {\tilde F}^n={ F^n \over \zeta}
~,~\,
\end{eqnarray}
\begin{eqnarray} 
{\tilde F}^S={ F^S \over\displaystyle {S+\bar S}}
~,~
{\tilde F}^T={ F^T \over\displaystyle {T+\bar T}}
~,~\,
\end{eqnarray}
and $K_{5, n} =\partial K_5/\partial Z_n$,
$K_{5, n \bar m} =\partial^2 K_5/(\partial Z_n \partial Z_{\bar m})$,
$\zeta_n =\partial \zeta/\partial Z_n$,
$\zeta_{n \bar m} =\partial^2 \zeta/(\partial Z_n \partial Z_{\bar m})$.

The original calculation 
is at order of $\epsilon$. Therefore, to be consistent, we should 
keep the scale, gauge couplings,
and  soft terms at order of $\epsilon$, because there may exist other
high order correction ( $\epsilon^2$ and higher ) to the
Lagrangian, or the K\"ahler potential and gauge kinetic function.
Considering $\epsilon^n = 0$ for n $>$ 1, 
 the following  scale and gauge coupling equations
 need to be changed:
\begin{eqnarray}
\alpha_{\rm GUT} &=&{{1-e} \over\displaystyle
 {2 V_p }}\,(4\pi\kappa^2 
)^{2/3} ~,~ \,
\end{eqnarray}
\begin{eqnarray}
\alpha_H &=&{{1-e_h} \over\displaystyle
 {2 V_p }}\,(4\pi\kappa^2 
)^{2/3} ~,~\,
\end{eqnarray} 
\begin{eqnarray}
M_H   = (1+e/6-e_h/6) M_{GUT} ~,~\,
\end{eqnarray}
\begin{eqnarray}
\left[\pi \rho_p\right]^{-1} &=& 
8 \pi \delta_{5b} (1- e) \left(2 
\alpha_{\rm GUT}\right)^{-3/2}
{{M_{GUT}^3}\over\displaystyle
  {M_{Pl}^2}}~,~\, 
\end{eqnarray} 
and the soft  terms at order $\epsilon$ are:
\begin{eqnarray} 
M_{1/2} = (1-e) {\tilde F}^S + e {\tilde F}^T
+ e {\tilde F}^n \zeta_n 
~,~\,
\end{eqnarray}
\begin{eqnarray}
M_0^2 &=& V_0 + M_{3/2}^2 -
\left( {2\over 3}
e  |{\tilde F}^S|^2 
+  |{\tilde F}^T|^2 
-{2\over 3} e Re{\tilde F}^S {\bar {\tilde F}}^{\bar T}
\right.\nonumber\\&&\left. 
+ {1\over 3} e \zeta \zeta_{n \bar m}
 {\tilde F}^n {\bar {\tilde F}}^{\bar m}
-{2\over 3} e \zeta_{\bar n} Re{\tilde F}^S {\bar {\tilde F}}^{\bar n}  
+{2\over 3} e \zeta_{\bar n} Re{\tilde F}^T {\bar {\tilde F}}^{\bar n} 
\right) 
~,~\,
\end{eqnarray}
\begin{eqnarray}
A&=& - \left( (1-e) {\tilde F}^S
+  e {\tilde F}^T
\right.\nonumber\\&&\left. 
+( e \zeta_n -  \zeta K_{5, n} )   
 {\tilde F}^n  \right)
~.~\,
\end{eqnarray}
Obviously, if $K_{5, n}$=0, we obtain $M_{1/2}=-A$ in this
case, too.
For the scale and gauge couplings, because we consider small
$\epsilon$ or the small next order correction, the
discussions are similar to those in ~\cite{JUNJUN, HLYLZH,
 DGCMUNOZ} 
with small $x$.
And because too many parameters in the soft term
expressions, we will not do the numerical analysis here,
for we can just let $M_{1/2}, M_0, A$ as free parameters.

\subsection{Toy Compactification and its Phenomenology}
In this section, we will consider the toy compactification
in non-standard embedding. First, we consider the
case without five-brane, which is similar to the
case in subsetion 2.2. Therefore, we just write down the
K\"ahler potential, gauge kinetic function and superpotential:
\begin{eqnarray}
K &=& \hat K + \tilde K^O |C_O|^2 
+\tilde K^H |C_H|^2 ~,~ \,
\end{eqnarray}
\begin{eqnarray}
\hat K &=&  -\ln\,[S+\bar S]-3\ln\,[T+\bar T] ~,~\,
\end{eqnarray}
\begin{eqnarray}
\tilde K^O &=& {3\over\displaystyle {T+\bar T}}
\left(1+{{\epsilon \beta^0 (T+\bar T)}\over\displaystyle 
{S+\bar S}}\right)^2
\left(1-{{\epsilon \beta^0 (T+\bar T)}\over\displaystyle 
{S+\bar S}}\right)
 ~,~ \,
\end{eqnarray}
\begin{eqnarray}
\tilde K^H &=& {3\over\displaystyle {T+\bar T}}
\left(1+{{\epsilon \beta^0 (T+\bar T)}\over\displaystyle 
{S+\bar S}}\right)
\left(1-{{\epsilon \beta^0 (T+\bar T)}\over\displaystyle 
{S+\bar S}}\right)^2
 ~,~ \,
\end{eqnarray}

\begin{eqnarray}
f^O_{\alpha \beta} &=& S \left(1+{{\epsilon \beta^0
 T}\over\displaystyle 
{S}}\right)^3
 \delta_{\alpha \beta} ~,~\,
\end{eqnarray}
\begin{eqnarray}
Ref^H_{\alpha \beta} &=& S \left(1-{{\epsilon \beta^0
 T}\over\displaystyle 
{S}}\right)^3
 \delta_{\alpha \beta} ~,~\,
\end{eqnarray} 
\begin{eqnarray}
W_O= \left(1+{{\epsilon \beta^0 T}\over\displaystyle 
{S}}\right)^{3/2}
\left(1-{{\epsilon \beta^0 T}\over\displaystyle 
{S}}\right)^{3/2} k d_{x y z} C_O^x C_O^y C_O^z ~,~\,
\end{eqnarray}
\begin{eqnarray}
W_H= \left(1+{{\epsilon \beta^0 T}\over\displaystyle 
{S}}\right)^{3/2}
\left(1-{{\epsilon \beta^0 T}\over\displaystyle 
{S}}\right)^{3/2} k d_{x y z} C_H^x C_H^y C_H^z ~,~\,
\end{eqnarray}
\begin{eqnarray}
W_{np} &=& h~ exp\left(-{{8 \pi^2}\over\displaystyle C_2(G^H)}
S \left(1-{{\beta T}\over\displaystyle 
{S}}\right)^3
 \right) ~.~\,
\end{eqnarray} 
The soft terms in the observable sector are similar to
those in the subsection 2.2. The only difference between
here and subsection 2.2 is that in section 2.2, the 
gauge group in the hidden sector
 is $E_8$, but here the gauge group in the hidden sector
  is the subgroup
of $E_8$.

Now, we consider the case with five-brane. The next order metric
is the following~\cite{NSLOW}:
\begin{eqnarray}
  ds^2 &=& (1+ y_{5b})
    g_{\mu \nu}^{(0)} dx^\mu dx^\nu 
       + (1-y_{5b})     
       e^{2a} g_{AB}^{(0)} dx^Adx^B 
\nonumber\\&&       
       + 
      (1 - 2 y_{5b}) 
       e^{2c} (dx^{11})^2 ~.~\,
\end{eqnarray}
M-theory limit is:   $  y_{5b} = {1\over 2}$
or $y_{5b} = - 1$.
In order to keep $ g_{11, 11} $ positive, we require that
$  y_{5b} < {1\over 2}$, and in order to keep the signature of 
metric $g_{\mu \nu}$ invariant, we require that
 $y_{5b} > - 1$. In short, we obtain:
 $ -1 <  y_{5b} < {1\over 2}$. 
The physical Calabi-Yau manifold's volume is 
obvious non-zero at any point along
the eleventh dimension, and if one defined $V_p^{min}$ and
$V_p^{max}$ as the minimum and maximum physical 
Calabi-Yau manifold's volume along the eleventh dimension,
respectively,
one obtains that:
\begin{eqnarray}
V_p^{max} <  64 V_p^{min} 
~.~\,
\end{eqnarray} 

Using same ansatz in section 2.2, we obtain the
K\"ahler potential, gauge kinetic funtion and the superpotential:
\begin{eqnarray}
K &=& \hat K  +\tilde K^O |C_O|^2 
+\tilde K^H |C_H|^2 ~,~ \,
\end{eqnarray}
\begin{eqnarray}
\hat K &=&  -\ln\,[S+\bar S]-3\ln\,[T+\bar T] + K_5 ~,~\,
\end{eqnarray}
\begin{eqnarray}
\tilde K^O &=& {3\over\displaystyle {T+\bar T}}
\left(1+{{\epsilon (T+\bar T)}\over\displaystyle 
{S+\bar S}}
\sum_{m=0}^{N+1} \left(1-{1\over 2} (Z_m +\bar Z_m)\right)^2
\beta^m \right)^2
\nonumber\\&&
\left(1-{{\epsilon (T+\bar T)}\over\displaystyle 
{S+\bar S}}
\sum_{m=0}^{N+1} \left(1-{1\over 2} (Z_m +\bar Z_m)\right)^2
\beta^m \right)
 ~,~ \,
\end{eqnarray}

\begin{eqnarray}
\tilde K^H &=& {3\over\displaystyle {T+\bar T}}
\left(1+{{\epsilon (T+\bar T)}\over\displaystyle 
{S+\bar S}}
\sum_{m=0}^{N+1} \left({1\over 2} (Z_m +\bar Z_m)\right)^2
\beta^m \right)^2
\nonumber\\&&
\left(1-{{\epsilon (T+\bar T)}\over\displaystyle 
{S+\bar S}}
\sum_{m=0}^{N+1} \left({1\over 2} (Z_m +\bar Z_m)\right)^2
\beta^m \right)
 ~,~ \,
\end{eqnarray}
\begin{eqnarray}
f^O_{\alpha \beta} &=& S 
\left(1+{{\epsilon T}\over\displaystyle 
{S}}
\sum_{m=0}^{N+1} \left(1-Z_m \right)^2
\beta^m \right)^3
\delta_{\alpha \beta} ~,~\,
\end{eqnarray}
\begin{eqnarray}
f^H_{\alpha \beta} &=& S 
\left(1+{{\epsilon T}\over\displaystyle 
{S}}
\sum_{m=0}^{N+1} Z_m^2
\beta^m \right)^3
\delta_{\alpha \beta} ~,~\,
\end{eqnarray}
\begin{eqnarray}
W_{np} &=& h~ exp\left(-{{8 \pi^2}\over\displaystyle C_2(G^H)}
S \left(1+{{\epsilon T}\over\displaystyle 
{S}}
\sum_{m=0}^{N+1} Z_m^2
\beta^m \right)^3 \right)
 ~.~\,
\end{eqnarray}

With the assumption $< Z_n > = < \bar Z_n > $,
$< S > = < \bar S >$, and $< T > = < \bar T >$,
we discuss the scale and gauge couplings first,
\begin{eqnarray}
8\pi\,G_{N}^{(4)} &=& {\kappa^2 \over 
{2\pi \rho_p V_p \delta_{5b}^T}} ~,~ \, 
\end{eqnarray}
\begin{eqnarray}
\alpha_{\rm GUT} &=&{1\over {2 V_p (1+e^T)^3}}\,(4\pi\kappa^2 
)^{2/3} ~,~ \,
\end{eqnarray}
\begin{eqnarray}
\alpha_H &=&{1\over {2 V_p (1+ e_h^T)^3}}\,(4\pi\kappa^2 
)^{2/3} ~,~\,
\end{eqnarray}
where 
\begin{eqnarray}
e^T&=& {{\epsilon (T+\bar T)}\over\displaystyle 
{S+\bar S}}
\sum_{m=0}^{N+1} \left(1-{1\over 2} (Z_m+\bar Z_m) \right)^2
\beta^m 
~,~\,
\end{eqnarray} 
\begin{eqnarray}
e_h^T={{\epsilon (T+\bar T)} \over\displaystyle 
{S+\bar S}}
\sum_{m=0}^{N+1} \left({1\over 2} (Z_m+\bar Z_m)\right)^2
\beta^m 
~,~\,
\end{eqnarray} 
\begin{eqnarray}
\delta_{5b}^T &=& {{\int_{0}^{\pi \rho} \sqrt {1-2 y_{5b} } (1+y_{5b}
)^2 (1-y_{5b} )^3 dx^{11}}
\over\displaystyle {\int_{0}^{\pi \rho} \sqrt {1-2 y_{5b} } dx^{11} }}
~.~\,
\end{eqnarray} 
The GUT scale $M_{GUT}$ and  the hidden sector GUT scale
$M_H$ when 
the Calabi-Yau manifold is compactified are:
\begin{eqnarray}
M_{\rm GUT}^{-6} &=& V_p ( 1+ e^T)^3 ~,~\,
\end{eqnarray}
\begin{eqnarray}
M_H^{-6} &=& V_p ( 1+e^T_h)^3 ~,~\,
\end{eqnarray}
or we can express the $M_H$ as:
\begin{eqnarray}
M_H &=&  ({{1+e^T} 
\over\displaystyle {1+e^T_h}})^{1/2} M_{GUT} ~,~\,
\end{eqnarray}
\begin{eqnarray}
M_{11} &=& \left[2 (4\pi )^{-2/3}\,  
\, \alpha_{\rm GUT}\right]^{-1/6} M_{GUT}  ~.~\,
\end{eqnarray}
And the physical 
 scale of the eleventh dimension is:
\begin{eqnarray}
\left[\pi \rho_p\right]^{-1} &=& 
{{8 \pi \delta_{5b}^T }\over\displaystyle {(1+e^T)^3}} \left(2 
\alpha_{\rm GUT}\right)^{-3/2}
{{M_{GUT}^3}\over\displaystyle
  {M_{Pl}^2}}~.~\, 
\end{eqnarray}
In order to keep $g_{11, 11}$ and $g_{1, 1}$ positive, we require that
$  -1 < y_{5b} < {1\over 2}$. So, we obtain:
\begin{eqnarray}
-{1\over 2} < e^T < 1 ~,~
-{1\over 2} < e^T_h < 1 ~.~\,
\end{eqnarray}
And the bounds on $M_H$ and $\alpha_H$ are:
\begin{eqnarray}
{1 \over {64}} \alpha_{GUT} < \alpha_H < 64 \alpha_{GUT}
~,~\, 
\end{eqnarray}
\begin{eqnarray}
{1\over 2} M_{GUT} < M_H < 2 M_{GUT}
~.~\, 
\end{eqnarray}
In addition, 
the discussions of the
scale and gauge couplings are similar to those
in ~\cite{JUNJUN, HLYLZH, DGCMUNOZ}
 with small $x$ if we defined the 
 physical eleventh dimension length as
 $\pi \rho_p  \delta_{5b}^T$. So, we will not redo the discussions
here again. Furthermore, the $\delta_{5b}^T$ might be at about
1 because of the constraints on $\beta^i$ and the charge distributions.

In addition, we obtain the following nomalized soft terms:
\begin{eqnarray}
M_{1/2} &=& {1\over\displaystyle
{S+\bar S+ \epsilon (T+\bar T)
 \Omega }}
\left( F^S \left(1- 2 {{\epsilon (T+\bar T) \Omega}\over\displaystyle 
{S+\bar S}} \right)  
 \right.\nonumber\\&&\left.
+ 3 \epsilon F^T  \Omega
 - 3 \epsilon F^n (T + \bar T) \Omega_n \right)
~,~\,
\end{eqnarray}
\begin{eqnarray}
M_0^2 &=& M_{3/2}^2 + V_0^2 - |F^S|^2
\left({{{\tilde K}^O_{S \bar S}} \over\displaystyle {\tilde K}^O}
-{{{\tilde K}_S^O {\tilde K}_{\bar S}^O}   
\over\displaystyle {{\tilde K}^{O2}}}\right)
\nonumber\\&&
- |F^T|^2 \left({{{\tilde K}^O_{T \bar T}} \over\displaystyle {\tilde K}^O}
-{{{\tilde K}_T^O {\tilde K}_{\bar T}^O}   
\over\displaystyle {{\tilde K}^{O2}}}\right)
-{\bar F}^{\bar n}\left({{{\tilde K}^O_{m \bar n}} \over\displaystyle 
{\tilde K}^O}
-{{{\tilde K}_m^O {\tilde K}_{\bar n}^O}   
\over\displaystyle {{\tilde K}^{O2}}}\right)
F^m
\nonumber\\&&
-2 \left({{{\tilde K}^O_{S \bar T}} \over\displaystyle 
{\tilde K}^O}
-{{{\tilde K}_S^O {\tilde K}_{\bar T}^O}   
\over\displaystyle {{\tilde K}^{O2}}}\right)
Re({\bar F}^{\bar T} F^S)
-2 \left({{{\tilde K}^O_{S \bar n}} \over\displaystyle 
{\tilde K}^O}
-{{{\tilde K}_S^O {\tilde K}_{\bar n}^O}   
\over\displaystyle {{\tilde K}^{O2}}}\right)
Re({\bar F}^{\bar n} F^S)
\nonumber\\&&
-2 \left({{{\tilde K}^O_{T \bar n}} \over\displaystyle 
{\tilde K}^O}
-{{{\tilde K}_T^O {\tilde K}_{\bar n}^O}   
\over\displaystyle {{\tilde K}^{O2}}}\right)
Re({\bar F}^{\bar n} F^T)
~,~\,
\end{eqnarray}
\begin{eqnarray}
A&=& F^S \left( -{1 \over\displaystyle {S+\bar S}}
-{{3 {\tilde K}_S^O}\over\displaystyle { {\tilde K}^O}}
\right)
+ F^T \left( -{3 \over\displaystyle {T+\bar T}}
-{{3 {\tilde K}_T^O}\over\displaystyle { {\tilde K}^O}}
\right)  
\nonumber\\&&
+ F^n \left( K_{5, n}
-{{3 {\tilde K}_n^O}\over\displaystyle { {\tilde K}^O}}
\right)
~,~\,
\end{eqnarray}
where
\begin{eqnarray}
{\tilde K}^O_S &=&
3\left(-{{\epsilon \Omega}\over\displaystyle 
{(S+\bar S)^2}} + 2
{{\epsilon^2 (T+\bar T) \Omega^2}\over\displaystyle 
{(S+\bar S)^3}} 
 +3{{\epsilon^3 (T+\bar T)^2 \Omega^3}\over\displaystyle 
{(S+\bar S)^4}}
\right)
~,~\,
\end{eqnarray}
\begin{eqnarray}
{\tilde K}^O_{S \bar S} &=&
6\left(  {{\epsilon \Omega}\over\displaystyle 
{(S+\bar S)^3}} -3
{{\epsilon^2 (T+\bar T) \Omega^2}\over\displaystyle 
{(S+\bar S)^4}} 
 -6{{\epsilon^3 (T+\bar T)^2 \Omega^3}\over\displaystyle 
{(S+\bar S)^5}}
\right)
~,~\,
\end{eqnarray}
\begin{eqnarray}
{\tilde K}^O_{S \bar T} &=&
6\left( 
{{\epsilon^2  \Omega^2}\over\displaystyle 
{(S+\bar S)^3}} 
 +3{{\epsilon^3 (T+\bar T) \Omega^3}\over\displaystyle 
{(S+\bar S)^4}}
\right)
~,~\,
\end{eqnarray}
\begin{eqnarray}
{\tilde K}^O_{S \bar n} &=&
3\left({{\epsilon \Omega_n}\over\displaystyle 
{(S+\bar S)^2}} -4
{{\epsilon^2 (T+\bar T) \Omega \Omega_n}\over\displaystyle 
{(S+\bar S)^3}} 
 -9 {{\epsilon^3 (T+\bar T)^2 \Omega^2 \Omega_n}\over\displaystyle 
{(S+\bar S)^4}}
\right)
~,~\,
\end{eqnarray}
\begin{eqnarray}
{\tilde K}^O_T &=&
3\left(-{1\over\displaystyle {(T+\bar T)^2}}
-{{\epsilon^2 \Omega^2} \over\displaystyle 
{(S+\bar S)^2}}
-2 {{\epsilon^3 (T+\bar T) \Omega^3} \over\displaystyle 
{(S+\bar S)^3}}
\right)
~,~\,
\end{eqnarray}
\begin{eqnarray}
{\tilde K}^O_{T \bar T} &=&
6\left({1\over\displaystyle {(T+\bar T)^3}}
- {{\epsilon^3 \Omega^3} \over\displaystyle 
{(S+\bar S)^3}}
\right)
~,~\,
\end{eqnarray}
\begin{eqnarray}
{\tilde K}^O_{T \bar n} &=&
6\left(
{{\epsilon^2 \Omega \Omega_n} \over\displaystyle 
{(S+\bar S)^2}}
+3 {{\epsilon^3 (T+\bar T) \Omega^2 \Omega_n} \over\displaystyle 
{(S+\bar S)^3}}
\right)
~,~\,
\end{eqnarray}
\begin{eqnarray}
{\tilde K}^O_n &=&
3\left(-{{\epsilon \Omega_n}\over\displaystyle {S+\bar S}}
+ 2 {{\epsilon^2 (T+\bar T) \Omega \Omega_n}
\over\displaystyle {(S+\bar S)^2}}
+3 {{\epsilon^3 (T+\bar T)^2 \Omega^2 \Omega_n}
\over\displaystyle {(S+\bar S)^3}}
\right)
~,~\,
\end{eqnarray}
\begin{eqnarray}
{\tilde K}^O_{n \bar l} &=&
3\left({{\epsilon \beta^n}\over\displaystyle 
{2(S+\bar S)}}
- {{\epsilon^2 (T+\bar T) \Omega \beta^n}
\over\displaystyle {(S+\bar S)^2}}
-{3\over 2} {{\epsilon^3 (T+\bar T)^2 \Omega^2 \beta^n}
\over\displaystyle {(S+\bar S)^3}}
\right)\delta_{n l}
\nonumber\\&&
+ 6\left(
-  {{\epsilon^2 (T+\bar T) \Omega_n \Omega_l}
\over\displaystyle {(S+\bar S)^2}}
-3 {{\epsilon^3 (T+\bar T)^2 \Omega \Omega_n \Omega_l}
\over\displaystyle {(S+\bar S)^3}}
\right)
~,~\,
\end{eqnarray}
and
\begin{eqnarray}
\Omega = \sum_{m=0}^{N+1} \left(1-{1\over 2} (Z_m +\bar Z_m)\right)^2
\beta^m 
~,~\,
\end{eqnarray}
\begin{eqnarray}
\Omega_n = \left(1-{1\over 2} (Z_n +\bar Z_n)\right) \beta^n
~.~\,
\end{eqnarray}

 Because we introduce the new parameters $\beta^i$ where
 i=1, N if we included five branes, we have a lot of freedom
 in phenomenology discussions.
 We do not do the numerical analysis of the soft terms
 because we just have 
 three soft term parameters $M_{1/2}, M_0^2$, and $A$, and  we can make them
 as free parameters by varying $\beta^i$ and $\epsilon$.   

\section{The Physical Calabi-Yau Manifold's Volume}

In above toy compactifications, we notice that the physical
Calabi-Yau manifold's volume at any point
along the eleventh dimension can not be zero from the metric
directly. And we can argue that this may be the general result.
The next order metric can be written as
~\cite{Witten, LOD}:
\begin{eqnarray}
  ds^2 &=&  (1+ {\sqrt 2 \over\displaystyle 6} {\bf B} )
   g_{\mu \nu}^{(0)} dx^\mu dx^\nu 
       + 
\left((1- {\sqrt 2 \over\displaystyle 3} {\bf B}) 
        g_{AB}^{(0)} + \sqrt 2 {\bf B}_{AB}
\right)
        dx^Adx^B 
\nonumber\\&&        
        + (1- {\sqrt 2 \over\displaystyle 3} {\bf B})
         (dx^{11})^2 ~,~\,
\end{eqnarray}
where $\partial_{[I_1} {\bf B}_{I_2...I_7]} ={1\over 7} * G_{I_1...I_7}$, and
 ${\bf B}_{ a \bar b}$ is defined as following~\cite{LOD}: 
\begin{eqnarray} 
{\bf B}_{\mu \nu \rho \sigma a \bar b}=
\epsilon_{\mu \nu \rho \sigma} {\bf B}_{a \bar b}
~,~\,
\end{eqnarray}
and $\bf B=  {g^{(0)}}^{A B} {\bf B}_{A B}$.
Because the massive modes are suppressed, 
$ {\bf B}_{AB}$ can be written as linear function of the
massless modes~\cite{LOD}. So,
 ${\sqrt 2 \over\displaystyle 3} {\bf B} 
g_{AB}^{(0)} $ and $\sqrt 2 {\bf B}_{AB}$  
have the same sign. In addition,
the magnitude of ${\sqrt 2 \over\displaystyle 3} {\bf B} 
g_{AB}^{(0)} $  is larger than that of 
$\sqrt 2 {\bf B}_{AB}$, for example, in the toy compactification, 
the magnitude of $\sqrt 2 {\bf B}_{AB}$ is half of that
of ${\sqrt 2 \over\displaystyle 3} {\bf B} 
g_{AB}^{(0)} $ which can be easily proved at next order in general. 
Therefore,
in order to keep
$g_{11, 11}$ positive, 
we may not push the physical Calabi-Yau manifold's 
volume  to zero at any point along the eleventh dimension. 

\section{Conclusion}
In M-theory on $S^1/Z_2$, we point out
that to be consistant, we should keep the scale, gauge couplings
and soft terms at next 
order in the standard embedding and non-standard 
embedding, and obtain
the soft term relations $M_{1/2} = -A$,
$|{{M_{0}}/{M_{1/2}}}| \leq 
{1/{\sqrt 3}}$ in the
standard embedding and soft term
relation $M_{1/2}=-A$ in non-standard embedding 
with five branes and $K_{5,n} =0$. Furthermore,
 we construct a toy compactification model which 
includes higher order
terms in 4-dimensional Lagrangian in standard 
embedding, and discuss its scale, gauge couplings, soft terms,
and explicitly show that the higher order terms do
affect the scale, gauge couplings and especially the soft terms
if the next order correction was not small.
We also construct a toy
compactification model in non-standard embedding with five branes,
calculate its K\"ahler potential, gauge kinetic function,  and discuss
the scale, gauge couplings and
soft terms. And M-theory limit gives strong constraints on
the gauge coupling ( $\alpha_H$ ) and GUT scale ( $M_H$ ) 
in the hidden sector in these toy models. 
 Finally, we argue that, in general, we might not
push the physical Calabi-Yau manifold's volume to zero at
any point along the eleventh dimension. 

The phenomenology consequences from
the soft terms obtained in this paper and 
the multi moduli toy compactification
are under investigation.

\section*{Acknowledgments}
This research was supported in part by the U.S.~Department of Energy under
 Grant No.~DE-FG02-95ER40896 and in part by the University of Wisconsin 
 Research Committee with funds granted by the Wisconsin Alumni
  Research Foundation.

\begin{figure}
\centerline{\psfig{file=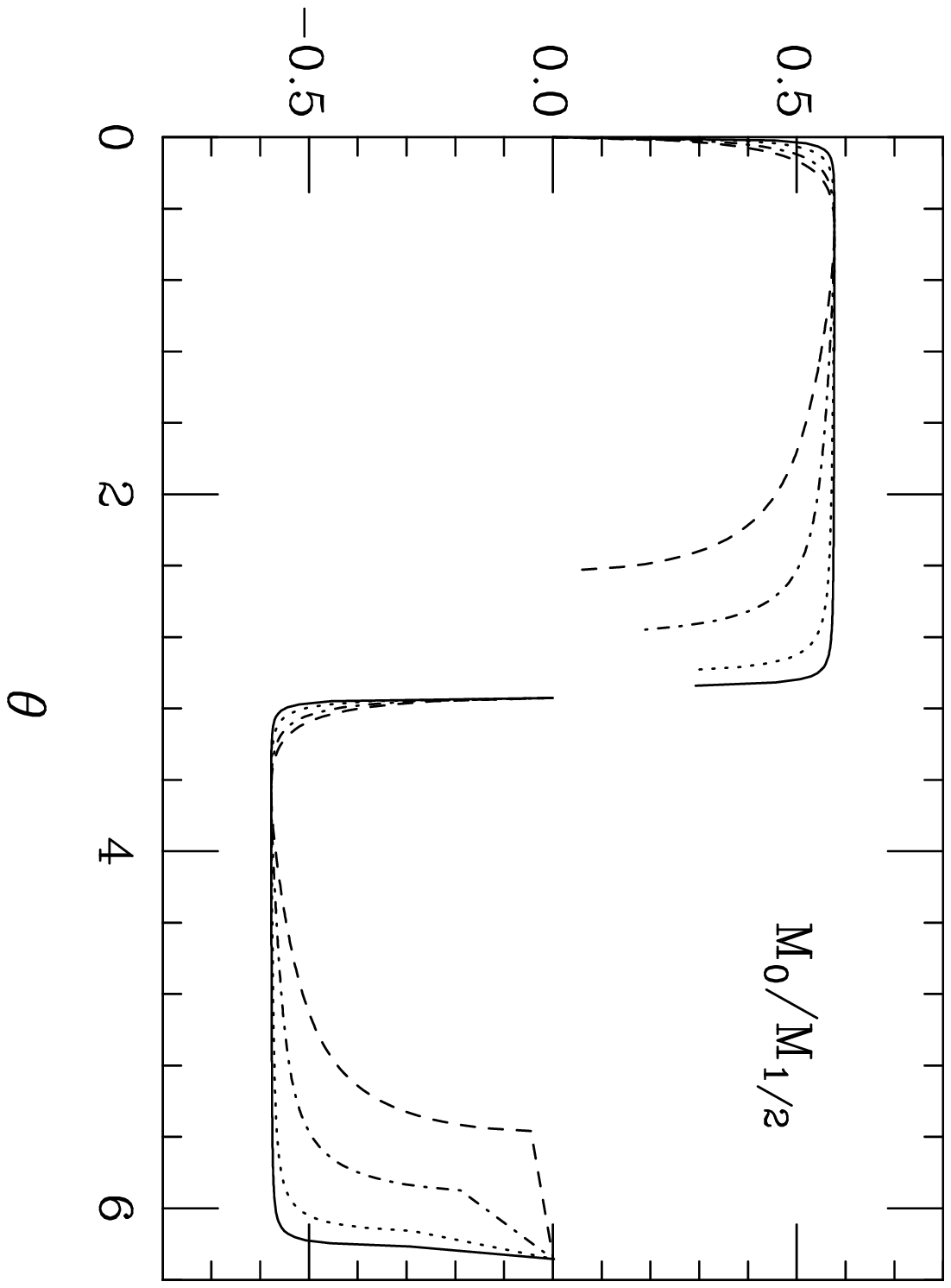,width=15cm}}
\bigskip
\caption[]{$M_0/M_{1/2}$ versus $\theta$.
 Solid line, dots line, dotdash, dashes line represent
x=0.05, 0.1, 0.2, 0.3, respectively.}
\label{diagrams}
\end{figure}

\begin{figure}
\centerline{\psfig{file=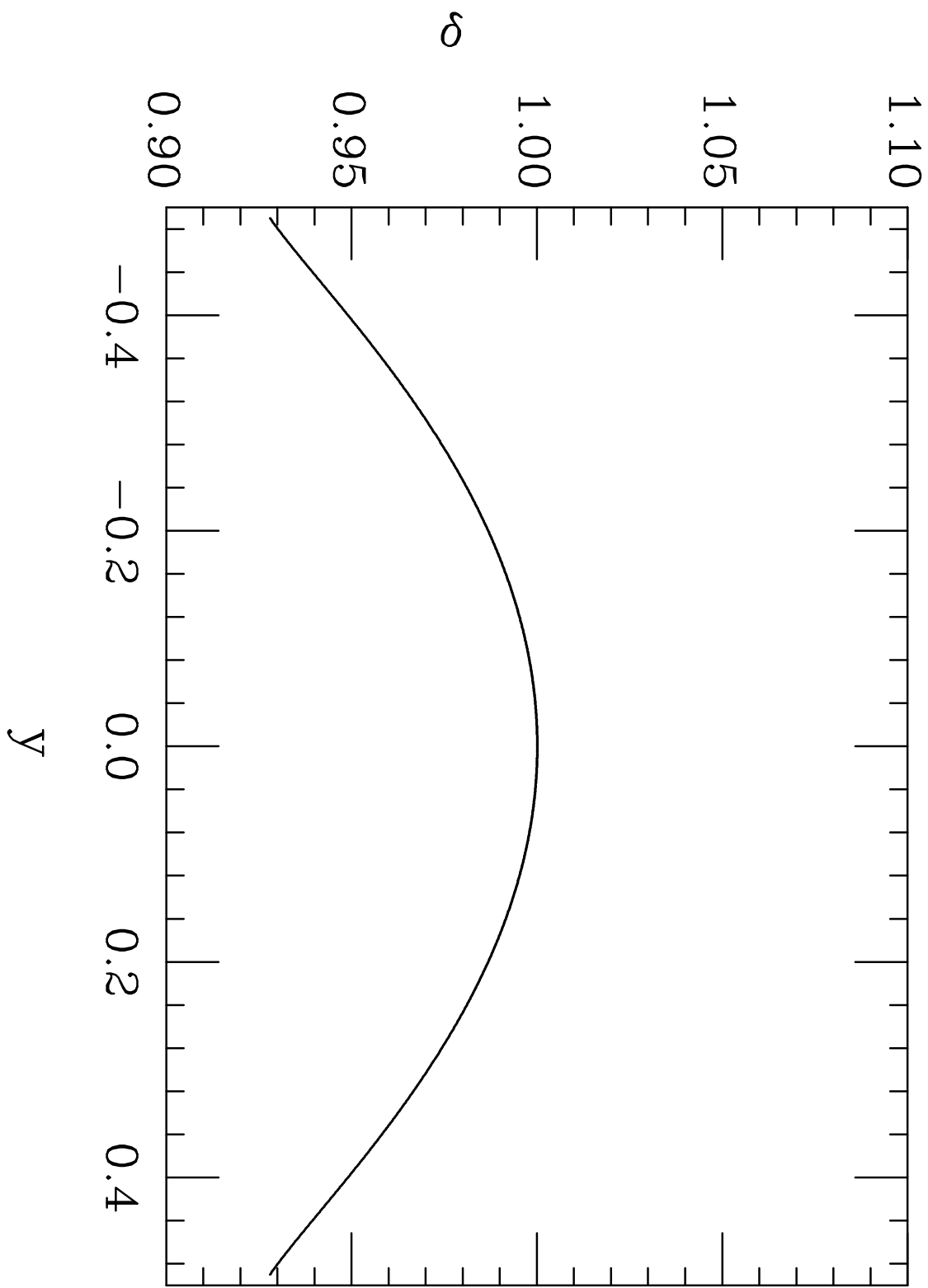,width=15cm}}
\bigskip
\caption[]{$\delta$ versus y.}
\label{diagrams}
\end{figure}

\begin{figure}
\centerline{\psfig{file=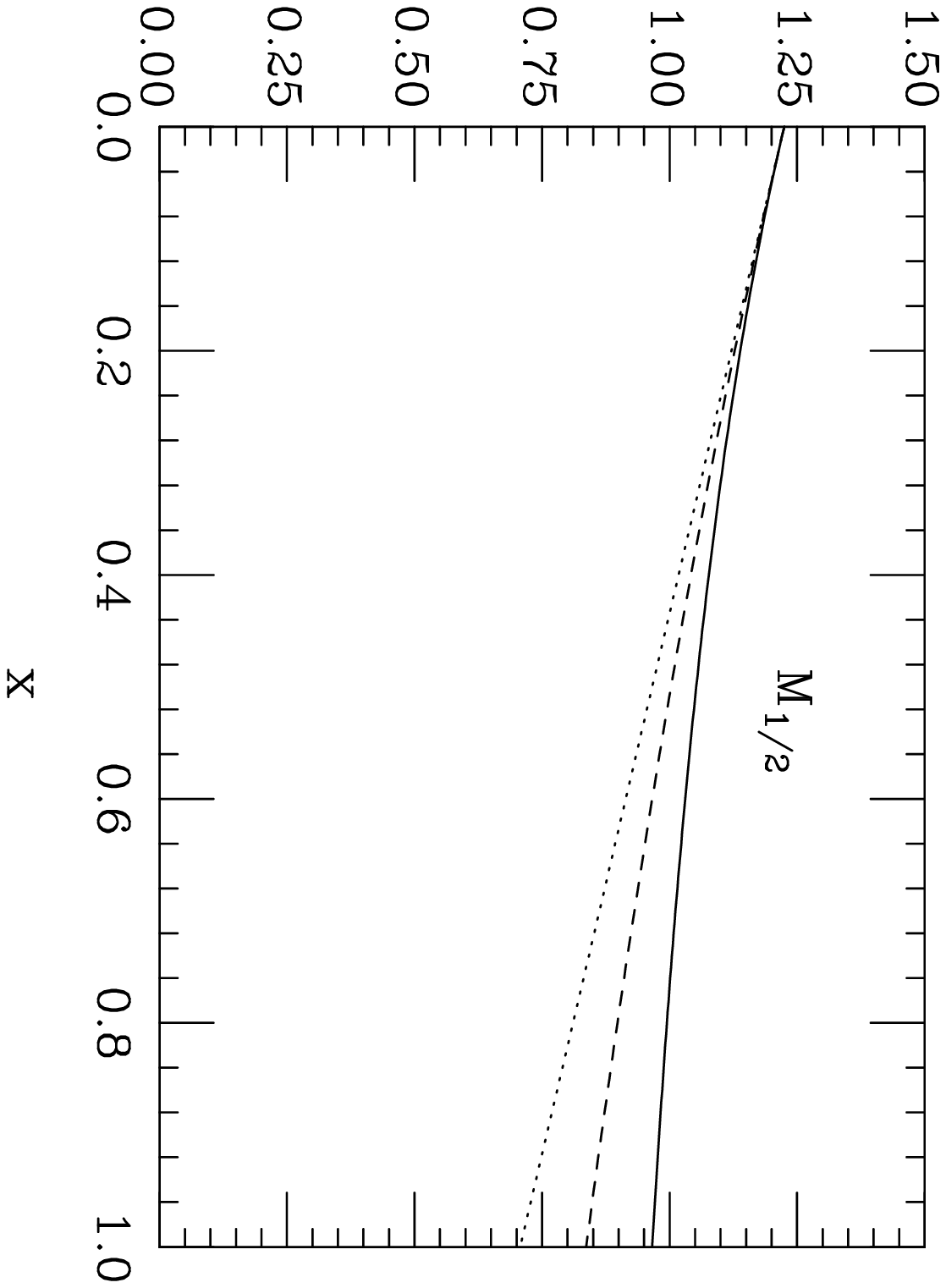,width=15cm}}
\bigskip
\caption[]{$M_{1/2}$ versus $x$  in
the unit of gravitino mass. Solid line, dots line, dashes line represent
scenario O, S, T, respectively.}
\label{diagrams}
\end{figure}

\begin{figure}
\centerline{\psfig{file=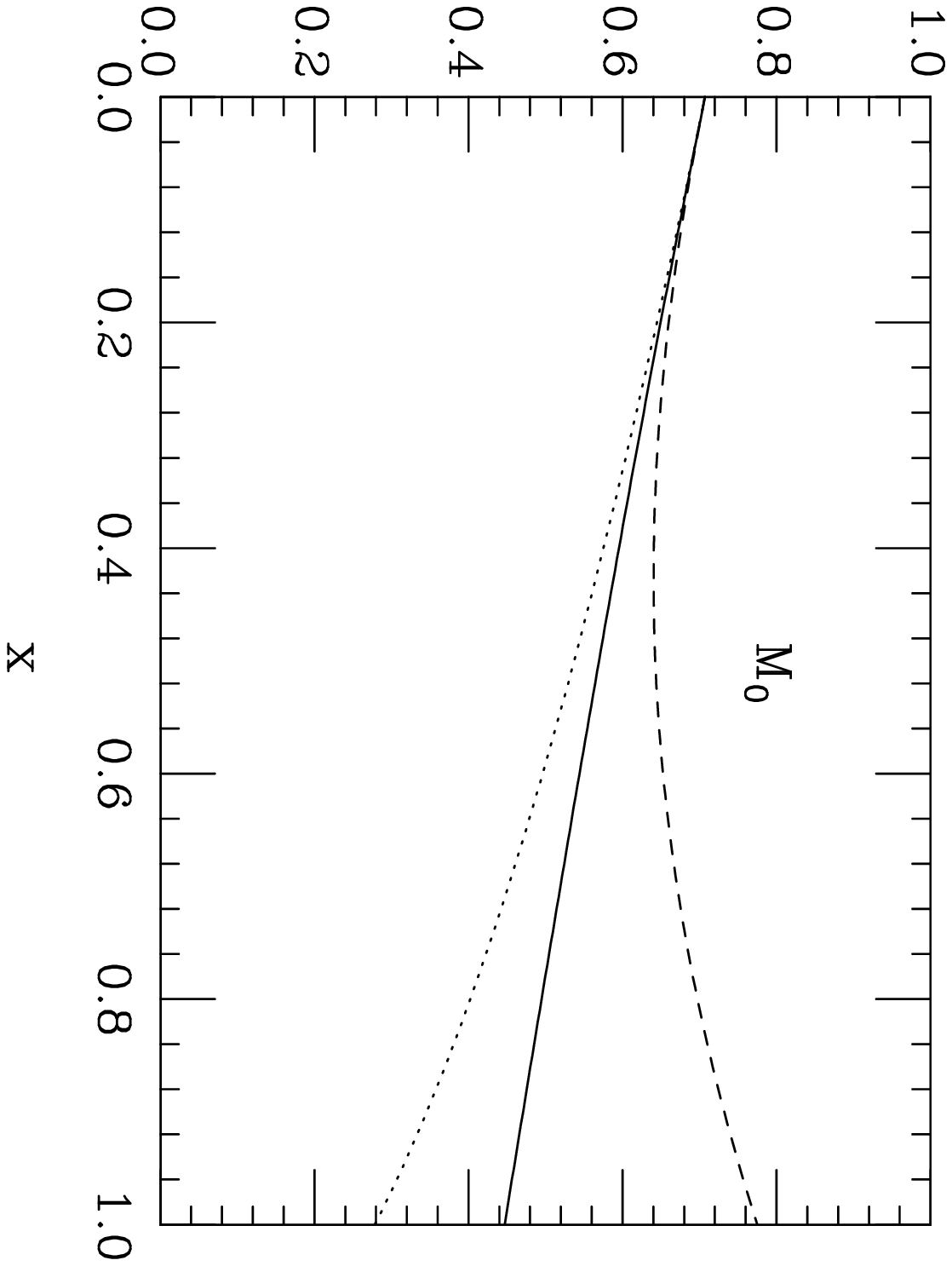,width=15cm}}
\bigskip
\caption[]{$M_0$ versus $x$  in
the unit of gravitino mass. Solid line, dots line, dashes line represent
scenario O, S, T, respectively.}
\label{diagrams}
\end{figure}

\begin{figure}
\centerline{\psfig{file=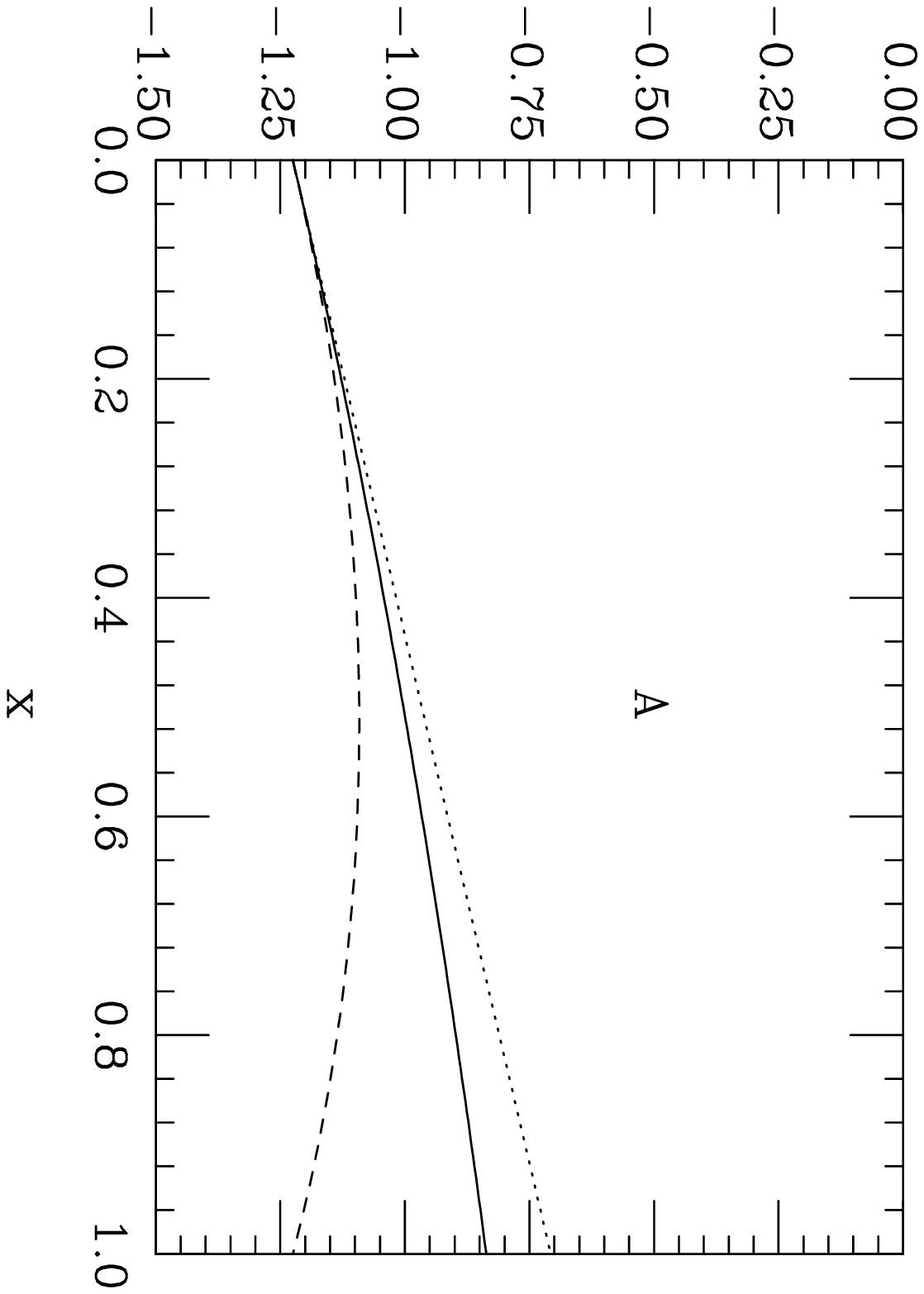,width=15cm}}
\bigskip
\caption[]{$A$ versus $x$  in
the unit of gravitino mass. Solid line, dots line, dashes line represent
scenario O, S, T, respectively.}
\label{diagrams}
\end{figure}

\begin{figure}
\centerline{\psfig{file=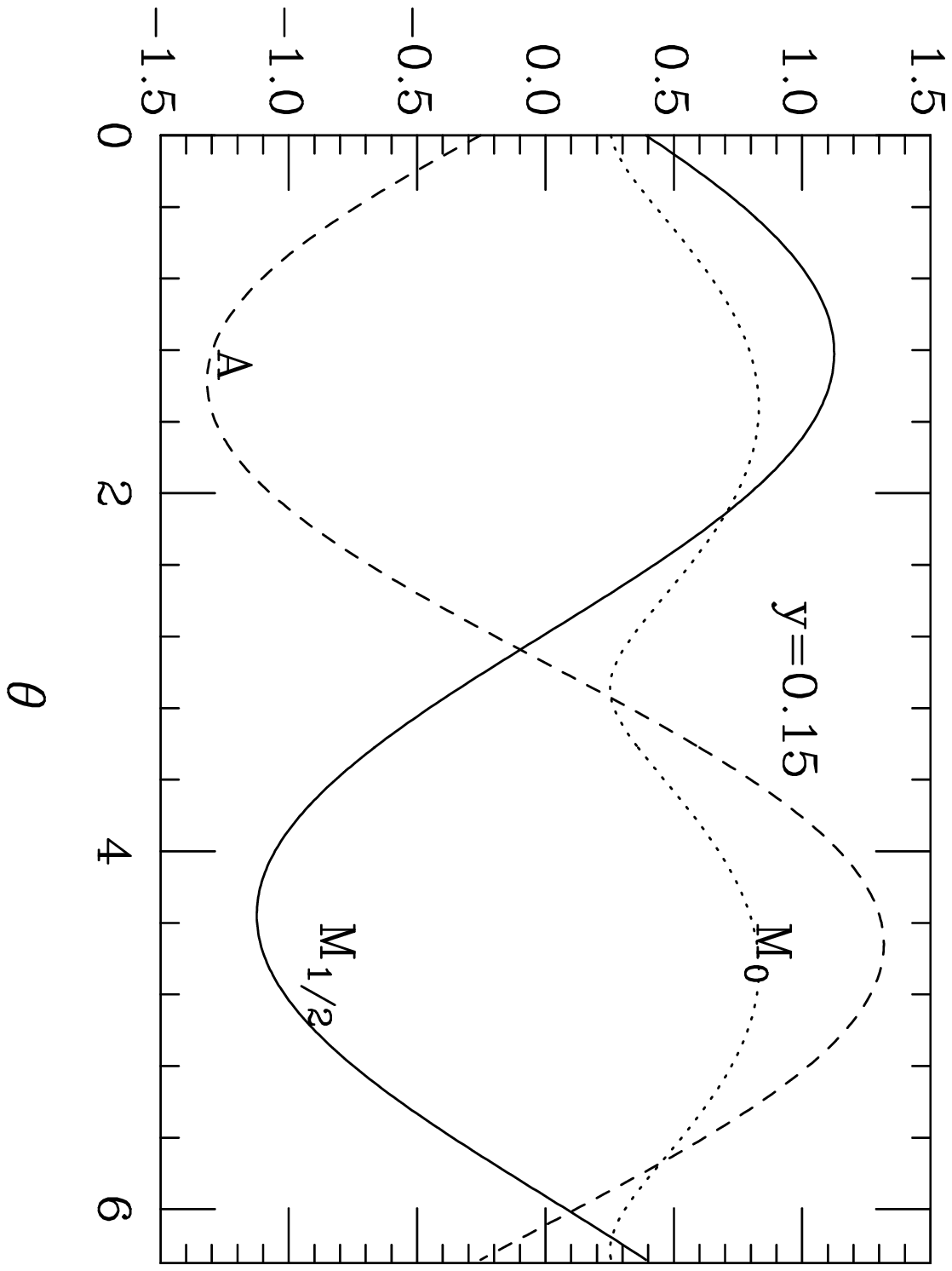,width=15cm}}
\bigskip
\caption[]{Soft terms versus angle $\theta$  with $y$=0.15 in
the unit of gravitino mass.}
\label{diagrams}
\end{figure}

\begin{figure}
\centerline{\psfig{file=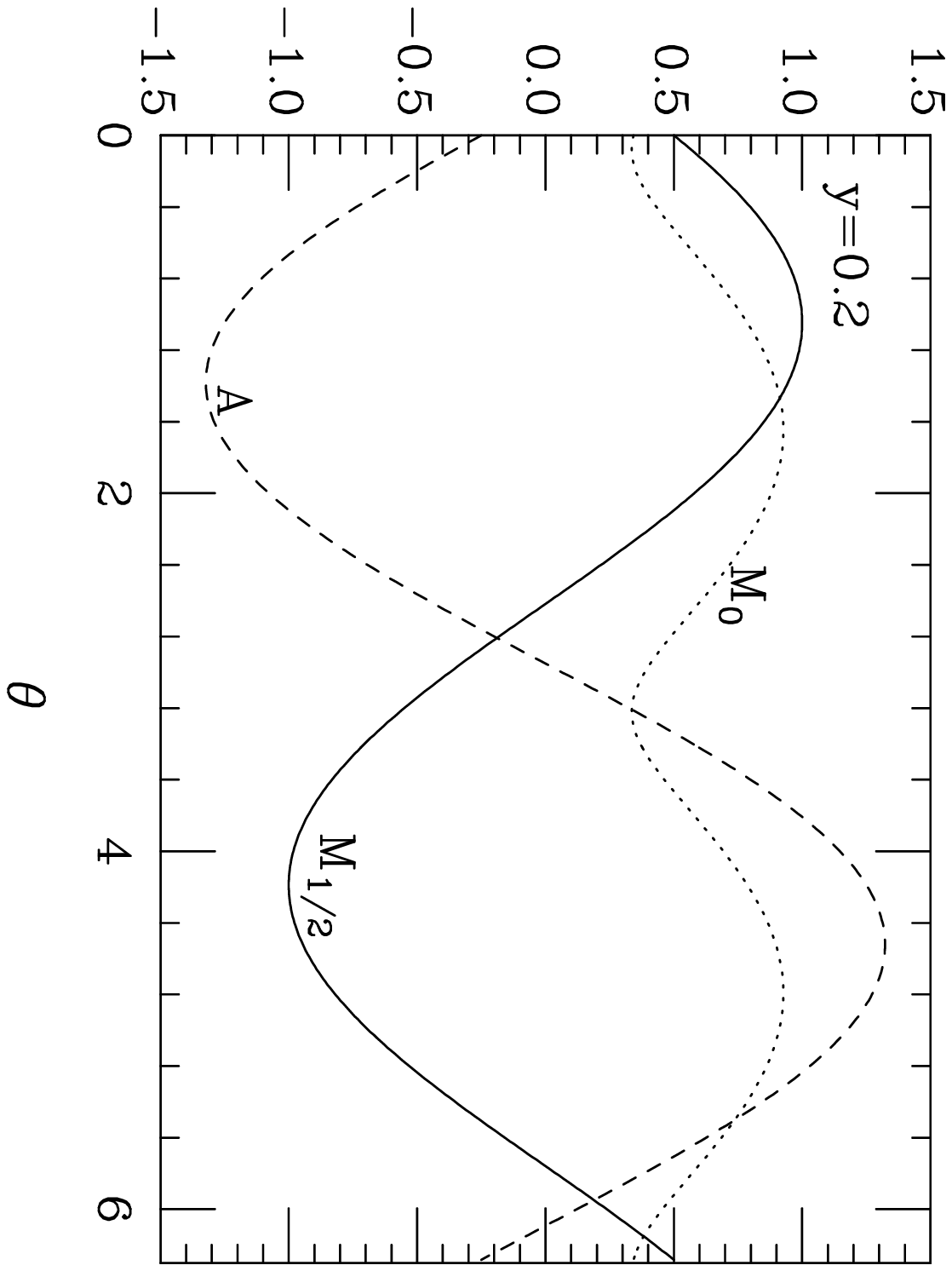,width=15cm}}
\bigskip
\caption[]{Soft terms versus angle $\theta$  with $y$=0.2 in
the unit of gravitino mass.}
\label{diagrams}
\end{figure}

\begin{figure}
\centerline{\psfig{file=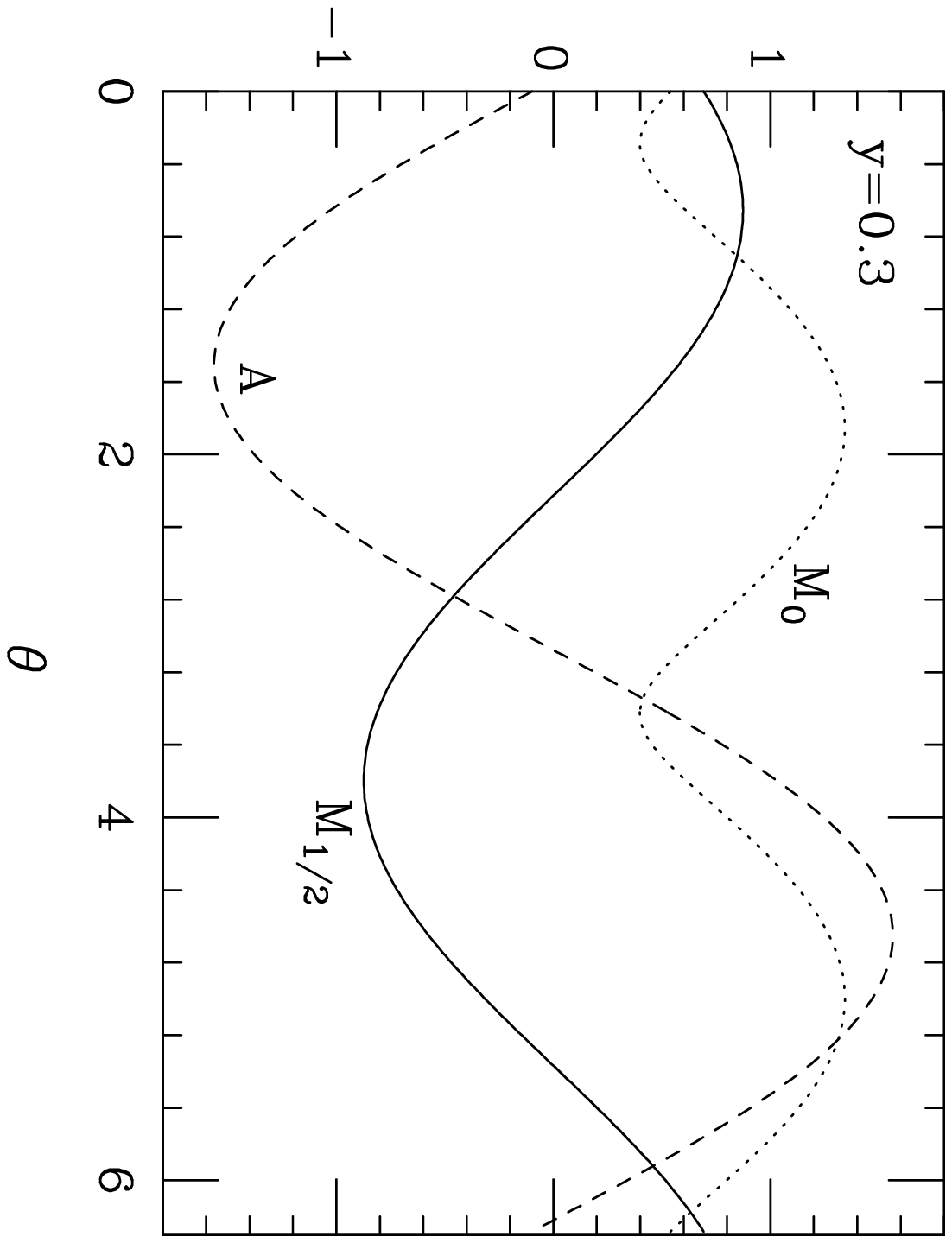,width=15cm}}
\bigskip
\caption[]{Soft terms versus angle $\theta$  with $y$=0.3 in
the unit of gravitino mass.}
\label{diagrams}
\end{figure}

\begin{figure}
\centerline{\psfig{file=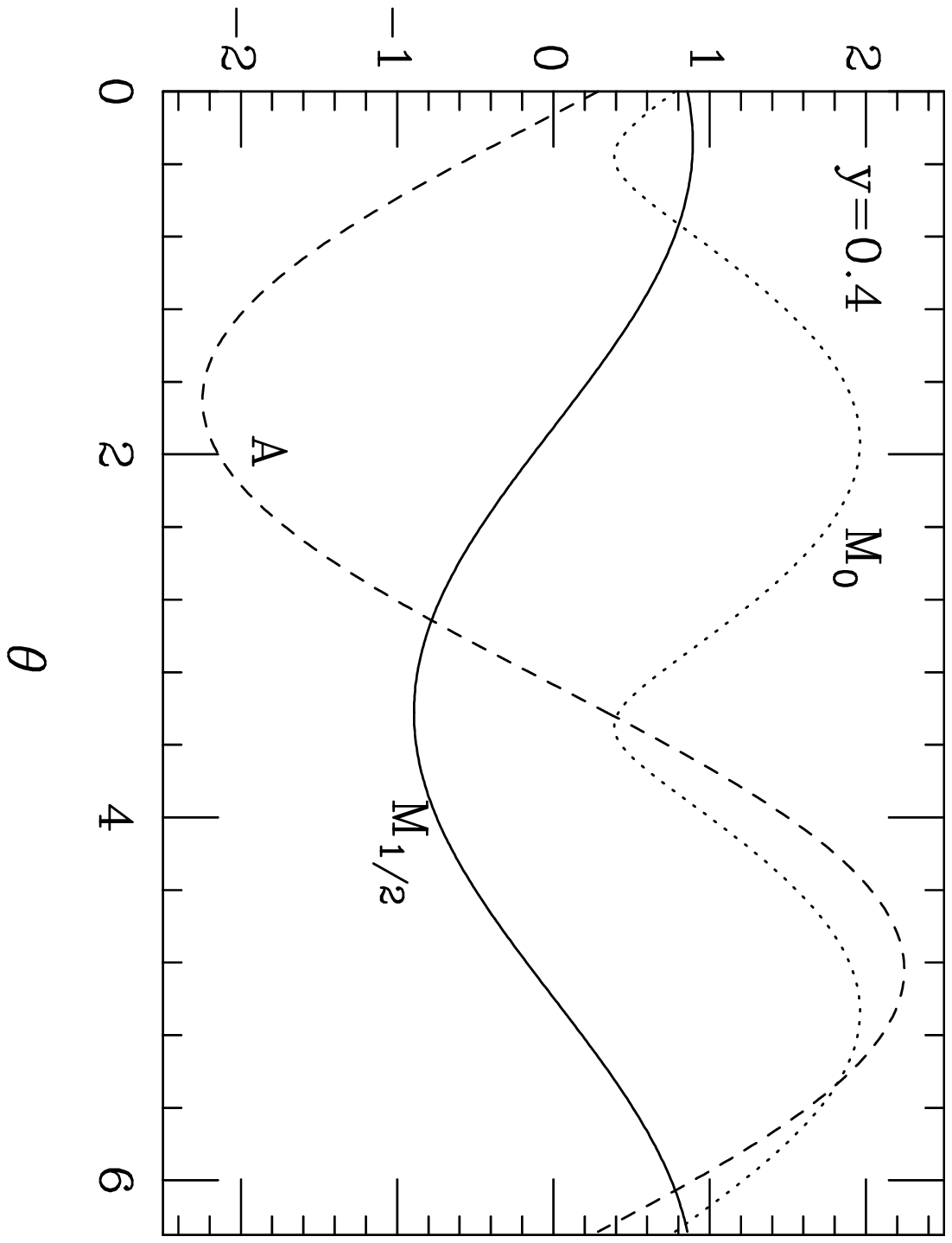,width=15cm}}
\bigskip
\caption[]{Soft terms versus angle $\theta$  with $y$=0.4 in
the unit of gravitino mass.}
\label{diagrams}
\end{figure}

\begin{figure}
\centerline{\psfig{file=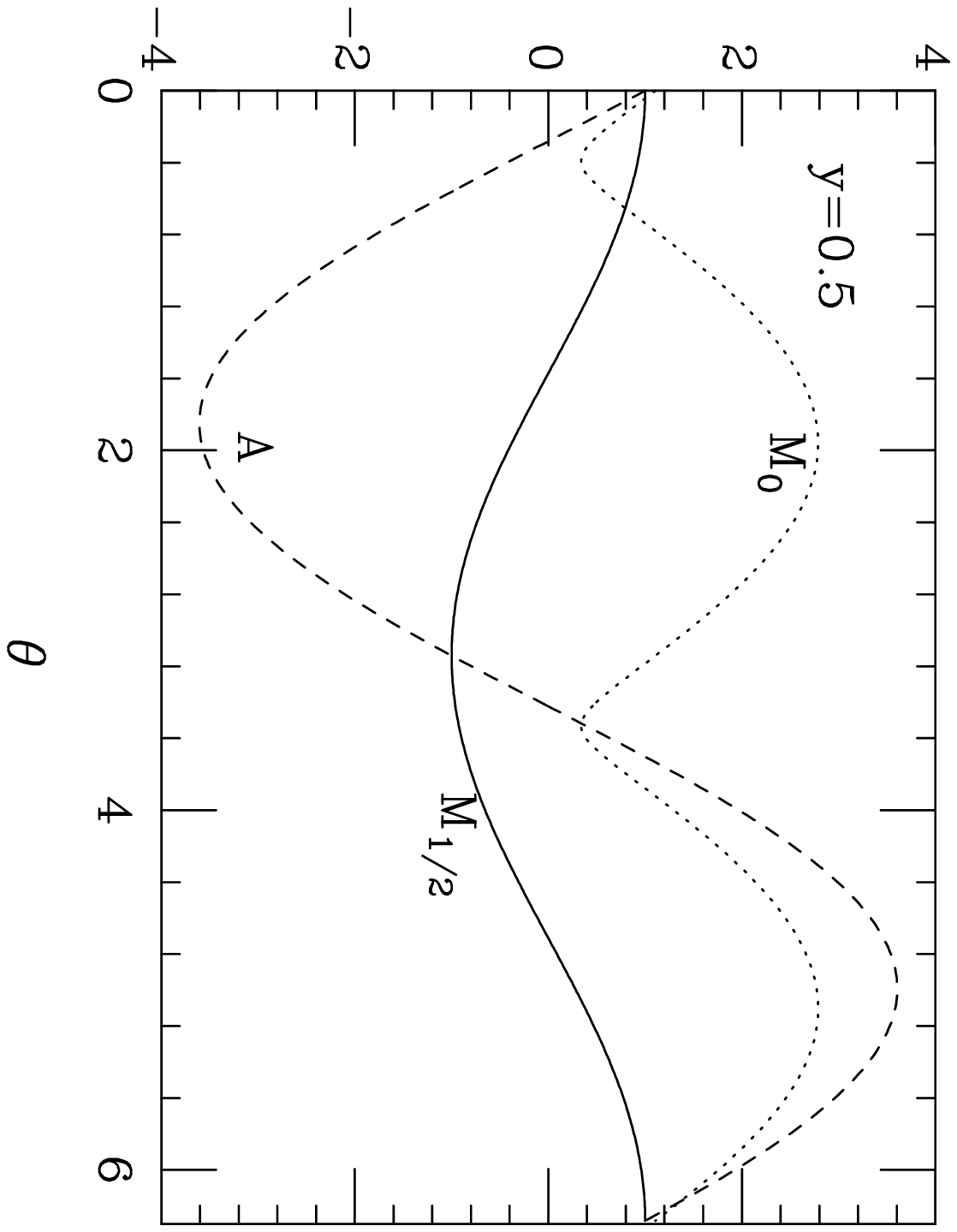,width=15cm}}
\bigskip
\caption[]{Soft terms versus angle $\theta$  with $y$=0.5 in
the unit of gravitino mass.}
\label{diagrams}
\end{figure}

\begin{figure}
\centerline{\psfig{file=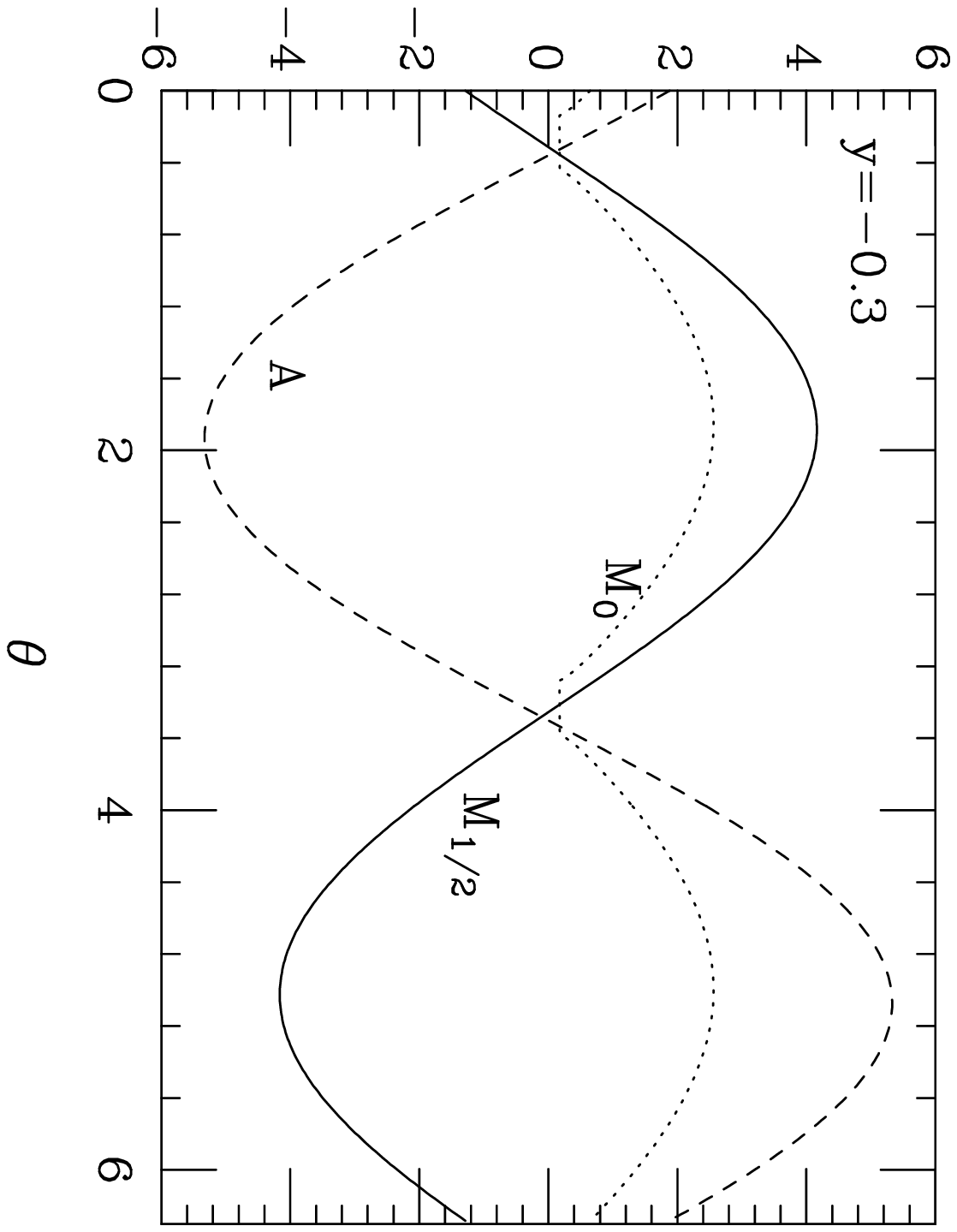,width=15cm}}
\bigskip
\caption[]{Soft terms versus angle $\theta$  with $y$=-0.3 in
the unit of gravitino mass.}
\label{diagrams}
\end{figure}

\end{document}